\documentclass[%
preprint,
showkeys,
bibnotes,
 amsmath,amssymb,
 aps,
]{revtex4-1}

\usepackage{graphicx}
\usepackage{dcolumn}
\usepackage{bm}

\usepackage{color}
\usepackage[dvipsnames]{xcolor}
\usepackage{soul} 
\usepackage{amsmath, amssymb}
\usepackage{epstopdf}
\usepackage{mathtools,euscript}
\usepackage{multirow}
\usepackage{mathrsfs, dsfont, yfonts}
\usepackage{lipsum}
\usepackage{calc}
\usepackage[french,english]{babel}
\usepackage{subfigure}
\usepackage{subfig}
\usepackage{floatrow}
\usepackage{svg}

\begin{document}

\preprint{APS/123-QED}

\title{Unveiling of the mechanisms of acoustic streaming induced by sharp edges}

\author{Chuanyu Zhang}
\affiliation{%
Universit\'e de Paris, LIED, UMR 8236, CNRS, F-75013, Paris, France.}%
  \email{chuanyu.dream@gmail.com}

\author{Xiaofeng Guo}%
\affiliation{%
Universit\'e de Paris, LIED, UMR 8236, CNRS, F-75013, Paris, France.}
\altaffiliation[Also at]{Universit\'e Gustave Eiffel, ESIEE Paris, F-93162, Noisy le Grand, FRANCE}
  \email{xiaofeng.guo@esiee.fr}

\author{Laurent Royon}
\affiliation{%
Universit\'e de Paris, LIED, UMR 8236, CNRS, F-75013, Paris, France.}%
  \email{laurent.royon@univ-paris-diderot.fr}

\author{Philippe Brunet}
\affiliation{Universit\'e de Paris, MSC, UMR 7057, CNRS, F-75013, Paris, France.}%
\email{philippe.brunet@univ-paris-diderot.fr}

\date{\today}

\begin{abstract}

Acoustic waves can generate steady streaming within a fluid owing to the generation of viscous boundary layers near walls, of typical thickness $\delta$. In microchannels, the acoustic wavelength $\lambda$ is adjusted to twice the channel width $w$ to ensure a resonance condition, which implies the use of MHz transducers. Recently though, intense acoustic streaming was generated by acoustic waves of a few kHz (hence with $\lambda \gg w$), owing to the presence of sharp-tipped structures of curvature radius at the tip $r_c$ smaller than $\delta$. The present study quantitatively investigates this sharp-edge acoustic streaming via the direct resolution of the full Navier-Stokes equation, using Finite Element Method. The influence of $\delta$, $r_c$ and viscosity $\nu$ on the acoustic streaming performance are quantified. Our results suggest choices of operating conditions and geometrical parameters, via dimensionless quantities $r_c/\delta$ and $\delta/w$ and provide guidelines on how to obtain strong, optimal sharp-edge acoustic streaming. 

\end{abstract}

\maketitle

\section{Introduction}
Acoustic streaming (AS) is a time-averaged steady flow generated by an acoustic field in a fluid, due to second-order nonlinear effects originating from the coupling between acoustics and hydrodynamics. The phenomenon has attracted researcher's attention for almost two centuries, since as early as 1831 when Faraday \cite{Faraday1831} first observed steady patterns of light particles on vibrating plates. More recently, AS has been proven to be a useful and non-invasive solution in various applied situations \cite{Friend2011}, like mixing under low-Reynolds number laminar flow conditions \cite{Sritharan2006}, particles manipulation and sorting \cite{Franke2010,Lenshof2012,Sadhal2012a,Muller2013,Skov2019,Qiu2019}, particles patterning \cite{Voth2002,Vuillermet2016} or heat transfer \cite{Legay2012,Loh2002}.

The underlying mechanism of AS lies in the dissipation of acoustic energy within a fluid induces spatial gradient of momentum, which creates a time-averaged effective forcing \cite{Westervelt1953,Nyborg1953,Lighthill1978,Friend2011,Eckart1948,Rayleigh1884,Schlichting,Nyborg1958,Riley1998a,Rayleigh2013}. Meanwhile, depending on the location of the main acoustic attenuation, AS can be induced either by viscous bulk fluid attenuation - denoted as Eckart streaming \cite{Eckart1948,Nyborg1953}), or by boundary layer attenuation - denoted as Rayleigh-Schlichting streaming \cite{Rayleigh1884,Schlichting,Nyborg1958,Riley1998a,Friend2011,Rayleigh2013}). 
For the latter, the development of an unsteady viscous boundary layer (VBL) along walls can lead to non-zero time-averaged Reynolds stress within this layer \cite{Schlichting}. Rayleigh's theory \cite{Rayleigh1884,Schlichting,Rayleigh2013} describes that the intense vorticity generated within the VBL appears as an array of eddies pairs (called inner vortices) aligned along the channel walls \cite{Andrade1931,Muller2013,Valverde2015}. This stress extends its influence beyond the VBL of thickness $\delta = \left(\frac{2 \nu}{\omega}\right)^{\frac{1}{2}}$ from the wall, 
and induces larger-scale eddies of width $\lambda /2$ \cite{Andrade1931,Hamilton2002} in the fluid bulk. 

To achieve AS in microfluidics geometries, the channel width and the wavelength are generally adjusted to ensure a resonance condition, typically obtained when $w \simeq \lambda/2$ \cite{Wiklund2012a}. Given the sound velocity in water and the main usual liquids being roughly between 1000 and 1800 m/s, $f$ shall be of the order of a few MHz. Therefore, while typical cost-effective transducers and associated amplifiers are generally in a range of a few kHz to a few tens of kHz, they should in principle fail to generate AS in microchannels, as the acoustic field would then be homogeneous in space. Although a few studies could circumvent this limitation by tuning the excitation of immersed bubbles \cite{Ahmed2009}, using micropillar \cite{Lux2019}, or flexural waves on a flexible wall \cite{Loh2002}, by prescribing a wavy channel geometry \cite{Lei2018,Subbotin2019,Jannesar2019}, or by tuning streaming modes within the transducer plane \cite{Lei2017b}, the majority of them were carried out under ideal situations such as infinite or semi-infinite domains and simple geometries. Still, remaining issues concern the influence of geometry, for instance the presence of obstacles or non-straight profiles like constrictions, or a situation of confinement when $\delta$ can be comparable to one of the channel dimensions \cite{Hamilton2002}.

-

\begin{table}
\caption{Definition of the main physical quantities}
\label{table:assembly}
\centering
\small
\renewcommand{\arraystretch}{1.25}
\begin{tabular}{l l}
\hline\hline
\multicolumn{1}{c}{Quantity} &
\multicolumn{1}{c}{Abbreviation} \\
\hline
Kinematic viscosity & $\nu$ \\
Viscous boundary layer thickness & $\delta$ \\
Tip angle of sharp edge & $\alpha$ \\
Height of the sharp edge & $h$ \\
Radius of curvature of the tip & $r_c$ \\
Width of the microchannel & $w$\\
Acoustic frequency & $f$  \\
Streaming velocity & $\textbf{v}_s$  \\
Acoustic vibration velocity & $\textbf{v}_a$\\
Orientation angle of the vibration velocity & $\alpha_{vb}$ \\
Maximum streaming velocity & $v_{sm}$\\
Maximum streaming velocity along the y-axis & $v_{sm}^{'}$ \\
Fitting coefficient relating $v_{sm}$ and $v_a^2$ & $\theta$ \\ 
Fitting coefficient relating $v_{sm}^{'}$ and $v_a^2$ & $\theta^{'}$ \\ 
\hline\hline
\end{tabular}
\normalsize
\end{table}

Recent studies have shown that intense AS could be generated via the coupling between acoustic waves and sharp structures \cite{Huang2013a,Huang2014,Nama2014,Nama2016a}. One of the particularities and main advantages of ``sharp-edge AS'' is that it is generated at relatively low frequency, typically in the kHz range. Meanwhile, the order of magnitude of the steady streaming velocity can even be comparable to the vibration velocity, hence up to several hundreds of mm/s \cite{Zhang2019}. Benefiting from this strong disturbance within the fluid, various applications using sharp structures streaming have been developed in microfluidics: mixing processes \cite{Huang2018a,Nama2014}, bio-particle control \cite{Leibacher2015,Cao2016}, as well as various on-chip devices \cite{Huang2014,Bachman2018}. 

However up to now, the underlying mechanisms of this streaming are not yet fully clear \cite{Ovchinnikov2014}. First, the pioneering study from T.J. Huang's group \cite{Huang2013a} attributes the induced streaming flow to the mechanical vibrations of the sharp structures induced by a transducer stuck on the microchannel wall. Such a vibration was indeed observed with high-speed imaging, and it raises the question on the adaptation of the sharp edge geometry to the prescribed frequency in order to ensure a resonance condition.
In Zhang \textit{et al.}'s study \cite{Zhang2019}, an oscillating flow was prescribed to the whole fluid, which also generates strong streaming around the sharp tip, but without the constraint of operating at a specific frequency.
Although Ovchinnikov \textit{et al'}s study \cite{Ovchinnikov2014} suggests that both situations should in principle lead to similar streaming flows, the first-order fluid oscillations should be different between the two situations.

Second, although both experiments \cite{Huang2013a,Zhang2019} and simulations \cite{Nama2014,Ovchinnikov2014} confirm the AS intensity depends on the sharpness of the tip, none of them dissociates the tip angle $\alpha$ from the curvature diameter $2r_c$, both of which being a sign of sharpness. The difficulty is that in practice, the micro-lithography techniques make these two quantities dependent on each other \cite{Zhang2019}. Therefore, only numerical simulations could help to tackle this challenging question. 
Third, while most studies on acoustic streaming generated around obstacles concern situations where $\delta \ll 2r_c$ and that of Ovchinnikov et al. \cite{Ovchinnikov2014} deals with the opposite situation ($\delta \gg 2r_c$), it is unclear how the crossover between the two situations takes place. Finally, from a theoretical point of view, sharp-edge AS remains a ground for a nonlinear framework in acoustofluidics equations. Indeed nonlinear terms coupling both the steady and periodic velocity fields can become dominant, or at least non-negligible, a feature which in turn makes the classical perturbation theory no longer adapted. This situation is the consequence of that, as mentioned above, the streaming velocity can be locally as strong as the vibration velocity \cite{Zhang2019}.

Motivated by these unsolved questions, and in the aim to propose quantitative predictions, the current study tries to address the AS flow under different operating conditions (vibration amplitude, sound frequency), fluid properties (viscosity) and geometries (tip sharpness quantified by both $r_c$ and $\alpha$). This parametric study is made possible by directly solving the full Navier-Stokes equation using Finite Elements Method.
Results from the DNS (Direct Numerical Simulation) are first validated by recent experiments, and then compared with those from simulations by classical Perturbation Theory (hereafter denoted as PT). This comparison points out the necessity to treat and include all non-linear terms in the numerical model. In a more applied purpose, this study aims to provide a framework for designing the optimal geometrical structure which would provide the strongest possible AS flow field. 

\section{Theoretical Model}
\label{theory} 
\subsection{Equations of motion}
The fundamental equations governing acoustic streaming have been previously presented in various theoretical studies \cite{Westervelt1953,Nyborg1958,Lighthill1978,Riley1998a,Friend2011,Bruus2012d,Sadhal2012}, which we summarize thereafter. Bold and normal font style respectively represent vectorial and scalar quantities. 
Without external body forces nor heat sources and for an isotropic homogeneous fluid, the mass and momentum conservation equations governing the flow are:
\begin{align} 
\frac{\partial \rho}{\partial t} + \bf{\nabla} \cdot (\rho \bf{v}) &= 0   \label{inc1} \\
 \rho \frac{\partial \bf{v}}{\partial t} + \rho (\bf{v} \cdot \nabla)\bf{v} &= \nabla\cdot\overline{\overline{\sigma}} \label{ns1}
\end{align}

\noindent where $\rho$ is the liquid density and $\bf{v}$ the velocity field. The Cauchy stress tensor $\overline{\overline{\sigma}}$ is the sum of the viscosity ($\mu$) term $\overline{\overline{\tau}}$ and pressure term $-p\overline{\overline{I}}$. As in our situation, $\lambda\gg w$, and that the Mach number $Ma=v_a/c \ll 1$, the fluid can be treated as being incompressible, leading to $\overline{\overline{\sigma}} = -p\overline{\overline{I}} + \mu (\nabla \bf{v} + {\nabla \bf{v}}^\intercal )$. Then, Eqs.~(\ref{inc1}) and (\ref{ns1}) can be reduced to:

\begin{align}
\nabla \cdot \bf{v}&=0 \label{inc2} \\
\rho \frac{\partial \bf{v}}{\partial t} + \rho (\bf{v} \cdot \nabla)\bf{v} + \frac{1}{\rho}\nabla p &=\mu \nabla^2 \bf{v} \label{ns2}
\end{align}

To analyse the AS flow, the Perturbation Theory (PT) constitutes the common general framework \cite{Westervelt1953,Nyborg1958,Lighthill1978,Riley1998a,Friend2011,Bruus2012d,Sadhal2012}. The velocity and pressure fields are decomposed into unperturbed state, oscillating and steady streaming parts, hereafter denoted with subscripts 0, $\omega$ and $s$, respectively:

\begin{subequations}\label{decomp}
\begin{align}
\textbf{v} =\textbf{v}_0+ \textbf{v}_{\omega}+{\textbf{v}}_{s},\ \textbf{v}_{\omega}=Re({\textbf{v}}_{a}e^{i\omega t})\\
p = p_0+p_{\omega}+p_{s},\ p_{\omega}=Re(p_{a}e^{i\omega t})
\end{align}
\end{subequations}

\noindent where ${\bf{v}}_0=\bf{0}$ is the unperturbed bulk flow considered to be null in this study, $\bf{v}_{\omega}$ is the acoustic (oscillating) part of the velocity field;
{$\bf{v}_{a}$ is the complex amplitude of the vibration velocity}, $\bf{v}_{s}$ is the steady streaming velocity; similarly, $p_{\omega}$,\ $p_{a}$ are the pressure and complex amplitude of the acoustic pressure field, $p_0$ is the gauge atmospheric pressure and $p_{s}$ is the steady pressure field associated to the streaming flow. The classical PT assumes $\left\lVert\textbf{v}_s\right\rVert \ll \left\lVert\textbf{v}_{a}\right\rVert$ and $p_s \ll p_{\omega}$, i.e. that the streaming flow velocity is of considerably lower magnitude than the driving acoustic velocity \cite{Nyborg1953,Boluriaan2003a,Friend2011,Bruus2012d,Sadhal2012,Ovchinnikov2014,Nama2014,Nama2016a,Lei2017}. Given the strong AS which is generated near sharp edges, we dismiss these simplifying assumptions.

By injecting the decomposition of Eq.~(\ref{decomp}) into Eqs. (\ref{inc2}) and (\ref{ns2}), and after a bit of algebra, the momentum equation leads to time-dependent (Eq.~\ref{nsosc}) and steady (Eq.~\ref{nssteady}) parts:

\begin{align}
& i\omega {v}_a + (\textbf{v}_s \cdot \nabla)\textbf{v}_a +(\textbf{v}_a \cdot \nabla)\textbf{v}_s =-\frac{1}{\rho}\nabla p_a +\nu \nabla^2 \textbf{v}_a \label{nsosc} \\
& (\textbf{v}_s \cdot \nabla)\textbf{v}_s +\frac{1}{2}Re[(\textbf{v}_a \cdot \nabla)\textbf{v}_{a}^*]=-\frac{1}{\rho}\nabla p_s +\nu \nabla^2 \textbf{v}_s\label{nssteady} 
\end{align}
 
Eqs.~(\ref{nsosc}) and (\ref{nssteady}) both contain non-linear terms in velocity, coupling the unsteady and steady components. By time-averaging Eq.~(\ref{nssteady}), one then sets a body force $\bf{F_s}$ to account for the non-linear effects of vibration motions \cite{Nyborg1953,Ovchinnikov2014}:

\begin{equation}
    (\textbf{v}_s \cdot \nabla)\textbf{v}_s = -\frac{1}{\rho} (\textbf{F}_s - \nabla p_s) +\nu \nabla^2 \textbf{v}_s
    \label{timeaveragedns}
\end{equation}

\noindent where the body force is:

\begin{equation}
    \textbf{F}_s = - \frac{\rho}{2} \left<  Re[(\textbf{v}_a \cdot \nabla)\textbf{v}_{a}^*] \right>
    \label{bodyforce}
\end{equation}

\noindent here the operator $<.>$ stands for a time-averaging over one period of acoustic oscillation $1/f$. 

In the PT framework, the non-linear terms at the left-hand side of eq. (\ref{nsosc}), coupling $\textbf{v}_a$ and $\textbf{v}_s$, are commonly neglected. Also in Eqs.~(\ref{nssteady}) and (\ref{timeaveragedns}), $(\textbf{v}_s \cdot \nabla)\textbf{v}_s$ is considered as a negligible, fourth-order term in most previous studies of acoustic streaming \cite{Nyborg1953,Boluriaan2003a,Friend2011,Bruus2012d,Sadhal2012,Ovchinnikov2014}. As stated above, in the case of sharp-edge streaming, ignoring these terms should deviate the modelled results from reality. The primary reason is, as previously mentioned, $\textbf{v}_{s}$ can be of the same order as $\textbf{v}_{a}$. It implies that the convection of the acoustic field by the streaming one becomes significant, as it was directly revealed by our previous experimental results, see inset of Figure 5 in \cite{Zhang2019}, especially in the upper range of acoustic velocity. The second reason lies in the boundary layer. Under usual situations where $\delta$ is much thinner than any other lengths of the problem - in particular, much smaller than the radius of curvature of the boundary walls, the resolution is carried out by solving separately the streaming flow within the steady VBL \cite{Nyborg1958,Boluriaan2003a,Sadhal2012} and that outside of the VBL. It consists of prescribing a distribution of slip velocities along walls, previously derived from the calculation within the VBL, to the fluid bulk. In the case of sharp edges when $r_c < \delta$, the direct numerical resolution in the whole domain, and especially within the VBL, becomes necessary. Ovchinnikov \textit{et al.}'s study \cite{Ovchinnikov2014} was dedicated to this situation, and our study is partly inspired by their approach. As our study aims to investigate streaming flows in an extended range of amplitude, we choose to keep these terms in our simulations.

\subsection{Qualitative view of the streaming force}

Let us now briefly examine the term $\textbf{F}_s$ of Eq. (\ref{bodyforce}). We assume that $\textbf{v}_a = \left[v_{ax} \, v_{ay} \, 0 \right] $ is a vector remaining in the $(xy)$ plane, which is true far from the upper and lower walls. Let us then calculate $\textbf{v}_a$ in this plane:

$$ (\textbf{v}_a \cdot \nabla)\textbf{v}_{a} = \left(\begin{array}{c} v_{ax} \dfrac{\partial v_{ax}}{\partial x} + v_{ay} \dfrac{\partial v_{ax}}{\partial y} \\ v_{ax} \dfrac{\partial v_{ay}}{\partial x} + v_{ay} \dfrac{\partial v_{ay}}{\partial y} \\  0 \\ \end{array}\right) $$

Results from our previously reported direct high-speed visualization \cite{Zhang2019} showed that, near sharp edges, the acoustic velocity field in fluid is aligned in parallel to the nearest wall. Furthermore, the no-slip boundary condition sets $\textbf{v}_a$=\textbf{0} along walls so that the amplitude of acoustic oscillations decreases to zero approaching the wall. This velocity gradient is the origin of shear stress within the VBL. 

In summary, gradients of acoustic velocity should originate from at least two effects: (i) the no-slip boundary condition which creates variation of velocity amplitude from $v_a = 0$ at the wall to $v_a \simeq A \omega$ at a distance to the wall farther than $\delta$, and (ii) the orientation of $\textbf{v}_a$ bending by an angle of $\pi - \alpha$ over a distance of $2 r_c$.

Along a straight horizontal wall, $v_{ay}$ is null and $v_{ax}$ is invariant with $x$. Therefore, only $v_{ax}$ and $\frac{\partial v_{ax}}{\partial y}$ take non-zero values, which implies that $\textbf{F}_s$ is null along a straight wall. This can easily be generalised along any straight wall of arbitrary orientation. However, the streaming force $\textbf{F}_s$ is non zero where there is a steep change of orientation of $\textbf{v}_a$, typically achieved near a sharp tip. This non-zero force is the origin of a centrifugal-like effect emphasised in previous studies \cite{Ovchinnikov2014,Zhang2019}. 

Let us finally remark that we deliberately choose to keep dimensional quantities in this study. First, our study aims for a quantitative comparison with previous experiments, which is made easier with dimensional quantities. Second, our problem involves four length scales which must be decoupled from each other.  More specifically, the acoustic wavelength ($\lambda$), the channel width (w), the VBL thickness ($\delta$) and the tip radius of curvature ($r_c$), must fulfil the condition: $\lambda \gg w \gg \delta \gg r_c$. This condition would lead to complex formulations for dimensionless equations. Thirdly, the COMSOL software we are using for our simulations, naturally works with dimensional quantities.

\section{Description of the numerical scheme}
Most of the numerical results presented in this paper are based on the direct solving of the Navier-Stokes equation (DNS). We also present a few results obtained from the Perturbation Theory (PT) inspired from Ovchinnikov \textit{et al.}'s study \cite{Ovchinnikov2014} as a matter of comparison between the two methods of their effectiveness under different conditions. Both PT and DNS simulations are conducted with Finite Element Method (FEM) using COMSOL Multiphysics \cite{Comsol}. 

Details of the simulation implementation techniques are described in Section \ref{appendix} (Appendix).

\subsection{Domain of study}

\begin{figure}
    \centering
    \includegraphics[width=\linewidth]{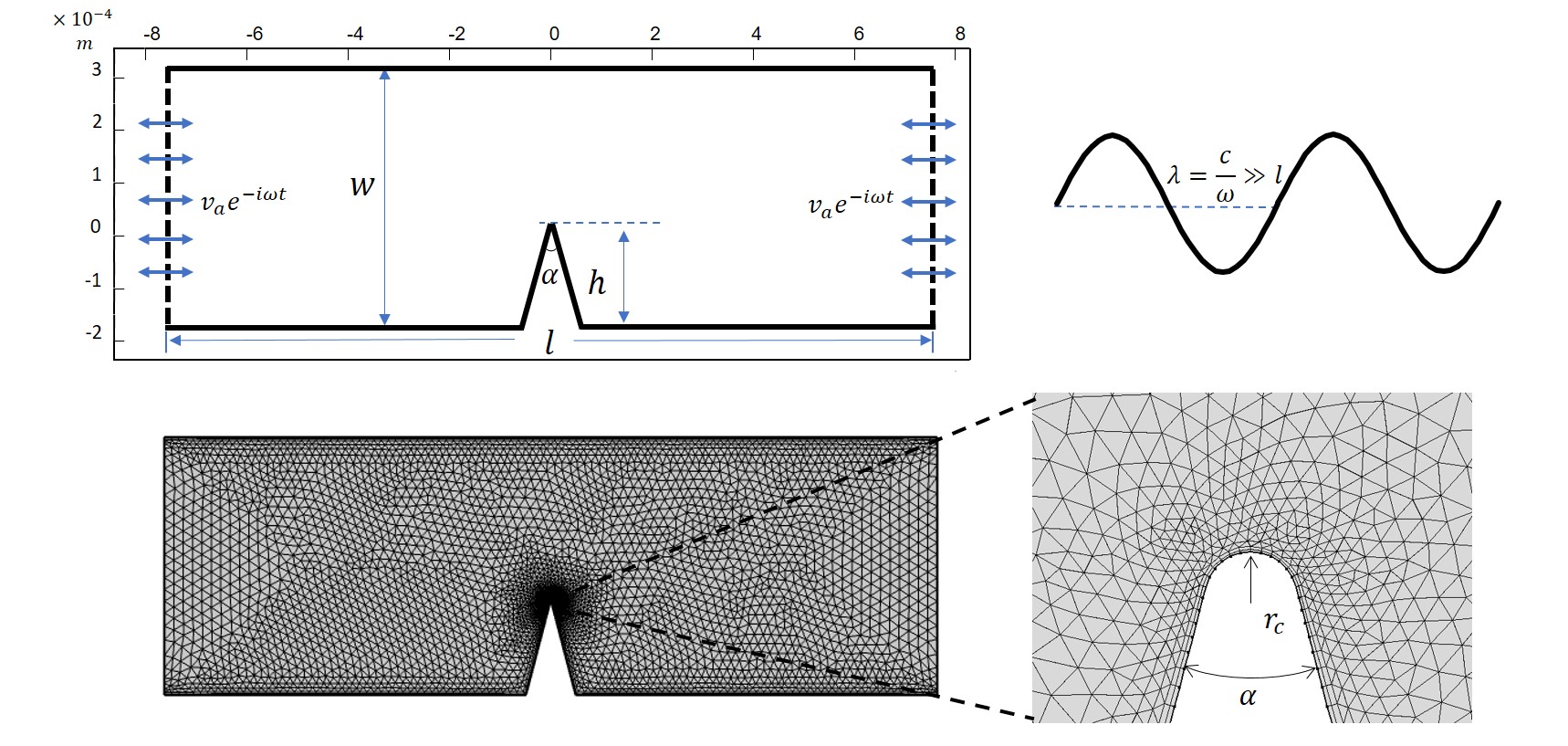}
    \caption{Geometry of the domain of study, sketched in the top-left inset, together with the acoustic wave parameters in the top-right inset. The mesh for the computation is shown in the bottom figure, with a magnified view around the sharp edge tip on the right.}
    \label{Fig_geometry_mesh}
\end{figure}

The geometrical dimensions of the microchannel with a sharp tip are detailed in Fig.~\ref{Fig_geometry_mesh}. Length and width of the channel are respectively $l$ = 1.5 mm and $w$ = 0.5 mm. A symmetrical sharp structure with a tip angle $\alpha$ and a curvature diameter $2r_c$ is located on one side of the channel. While both $\alpha$ and $2r_c$ are variable parameters for different simulation cases, the height of the sharp structure is kept constant: $h$=0.18 mm. 

It worth noting that the simulations are conducted in the framework of a bidimensional (2D) geometry. Precisely, the channel is considered infinitely deep. This choice is justified by two main reasons. First, all previous experiments of sharp-edge streaming including ours, are conducted with water and $f$ equal to a few kHz, yielding $\delta$ between 8 and 15 $\mu$m, while the channel depth $d$ is equal or larger than 50 $\mu$m. Second, the cross-sectional depth/width aspect ratio is roughly 1/10. As a consequence, the streaming develops essentially within the $(xy)$ plane. 

Near the sharp edge, the mesh is refined (Fig.~\ref{Fig_geometry_mesh}) since velocity gradients - thus the streaming force, are supposed to be locally concentrated near the tip. The mesh refining also allows to accurately account for the sharp geometry of the tip. Furthermore, a 3-layer inflation is created within the VBL of both channel and sharp-edge limits. This is essential to finely simulate the effect of viscous shear stress on acoustic streaming. 

\subsection{Boundary conditions}

Different from the PT method, DNS directly computes the fluid motion coupling equations for the acoustic oscillations (time-periodic) and for the steady streaming. Periodic boundary conditions are set at left and right ends of the channel. The left end (here set as the inlet) is attributed a periodic velocity $v_{x}= v_{a}sin(2\pi \omega t)$ along the horizontal direction and $v_{y}=0 $ along the vertical one. For the right end (outlet), a condition of pressure fixed at $p_0$ is assigned. 

Since the fluid remains incompressible, and the length scale of the domain is much smaller than the acoustic wavelength ($l \ll \lambda$), the above conditions result in an in-phase periodic velocity for the right and left borders, as shown in Fig.~\ref{Fig_geometry_mesh}. These conditions are supported by experimental observations of oscillations of fluid particles within the whole channel, while the sharp-edged tip remains static in the laboratory frame \cite{Zhang2019}. No-slip condition is prescribed on all other channel boundaries including the sharp edge itself.

For the time-dependent simulations, each acoustic period is discretized into 50 time steps, for an overall duration of 30 acoustic periods. It turns out that this duration is sufficient to allow the full establishment of a quasi-steady acoustic streaming, once the flow is averaged over an acoustic period. Moreover, the choice of 50 time steps per acoustic period is validated by comparing the streaming results from 4 different time steps. This validation process is documented in Section \ref{appendix} (Appendix) and shown in Fig.~\ref{figgridtimeinde}.

\section{Results and Discussions}
\subsection{Validation of the numerical scheme}
\begin{figure}
  \centering 
  \includegraphics[width=\linewidth]{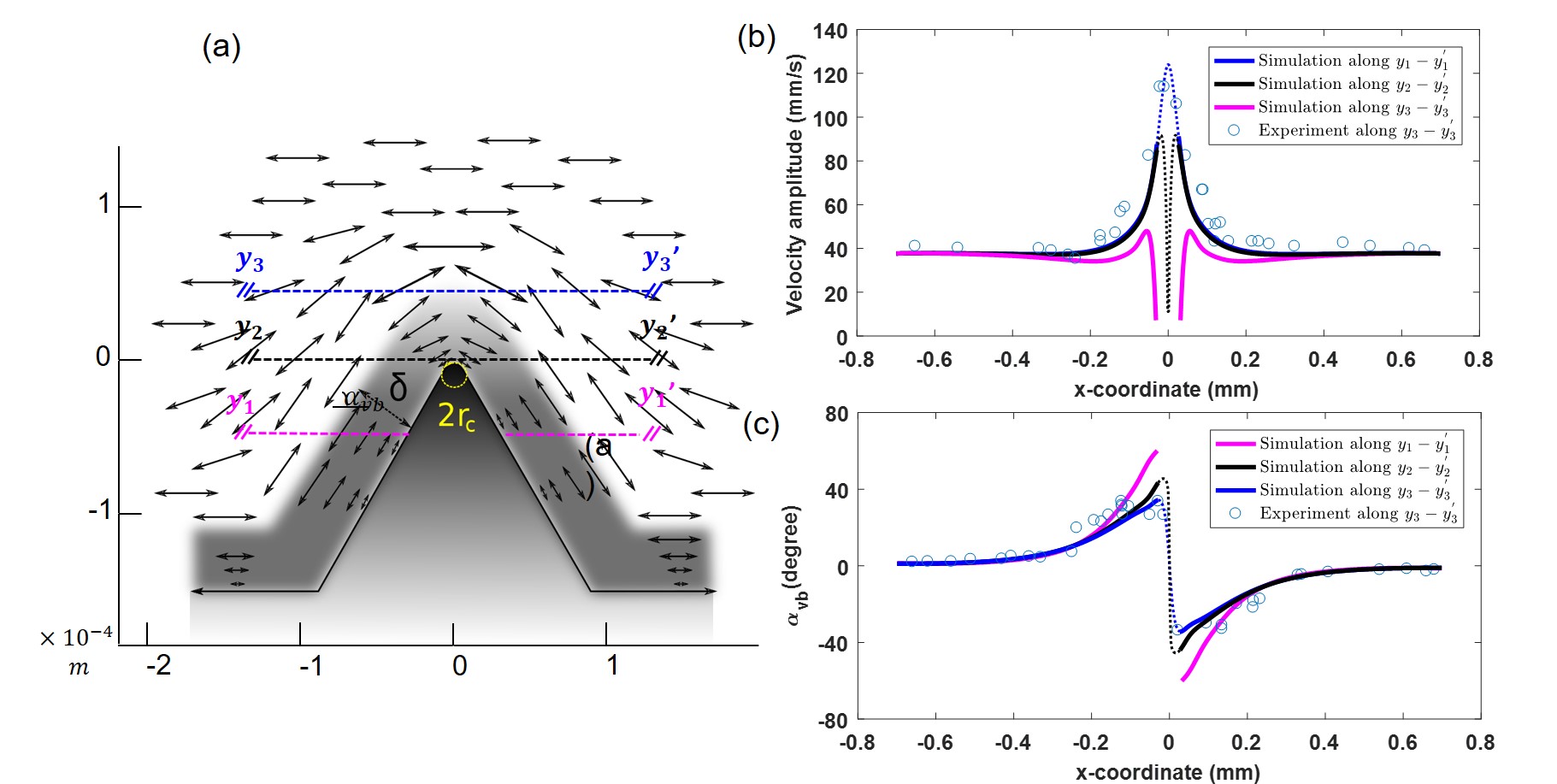}
  \caption{Acoustic vibration and streaming flow around the sharp edge structure: (a) Sketch of the acoustic vibrations of fluid particles near the sharp edge, $\delta$ is the acoustic boundary layer, the segment  $y_{1}-y_{1}^{\prime}$ is located 0.05 mm below the tip; $y_{2}-y_{2}^{\prime}$ intersects the tip; $y_{3}-y_{3}^{\prime}$ is located 0.01 mm above the tip, (b) Amplitude of the vibration velocity recorded along $y_{1}-y_{1}^{\prime}$, $y_{2}-y_{2}^{\prime}$ and $y_{3}-y_{3}^{\prime}$. Circles stand for experiments recorded along $y_{3}-y_{3}^{\prime}$, (c) Orientation of the vibration velocity $\alpha_{vb}=\arctan[v_{ay}/v_{ax}]$ along $y_{1}-y_{1}^{\prime}$, $y_{2}-y_{2}^{\prime}$ and $y_{3}-y_{3}^{\prime}$. Circles stand for experiments recorded along $y_{3}-y_{3}^{\prime}$.
  Parameters: $\alpha = 60^{\circ}$, $2r_c$=5.8 $\mu$m, $f$=2500 Hz, $v_a$=37.8 mm/s, $\delta$=11.5 $\mu$m.}
  \label{fig_vibration1}
\end{figure}

\begin{figure}
  \centering 
  \includegraphics[width=\linewidth]{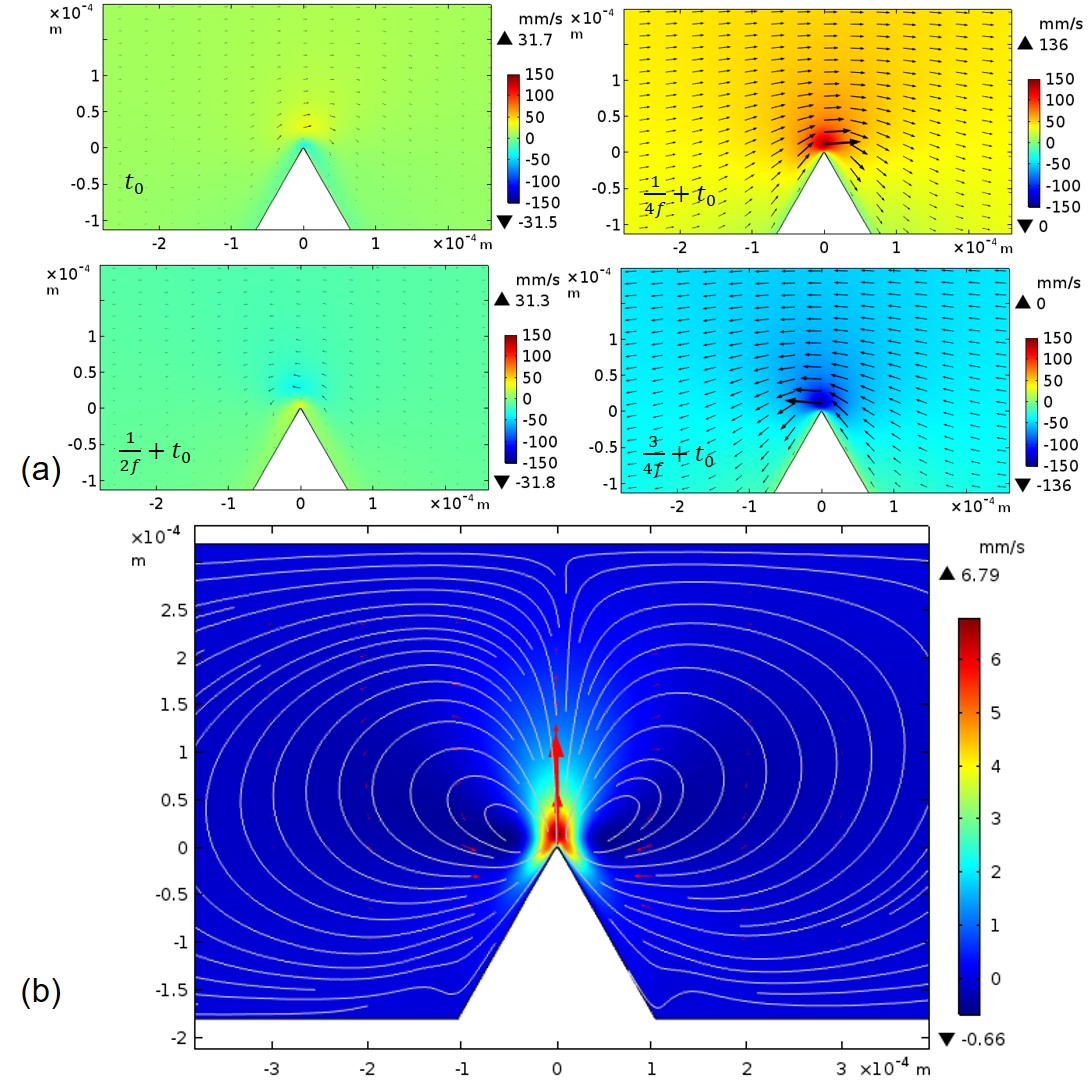}
  \caption{Acoustic vibration and streaming flow around the sharp edge structure: (a) Successive velocity fields at different time during one acoustic period, (b) Magnitude and streamlines of the streaming flow, from a time-average during several acoustic cycles.
  Parameters: $\alpha = 60^{\circ}$, $2r_c$=5.8 $\mu$m, $f$=2500 Hz, $v_a$=37.8 mm/s, $\delta$=11.5 $\mu$m.}
  \label{fig_vibration2}
\end{figure}

The comparison between previous experimental results \cite{Zhang2019} and present DNS ones, ensures the validation of our numerical scheme. Figures \ref{fig_vibration1} and \ref{fig_vibration2} intend to illustrate the mechanism of acoustic streaming, by showing both typical acoustic and steady velocity fields. Figure \ref{fig_vibration1}(a) presents a qualitative sketch and Figures \ref{fig_vibration1}-(b-c) show typical amplitude and orientation of the acoustic velocity field from both experiments and numerical simulations. 

\subsubsection{Acoustic velocity}

As shown in Fig.~\ref{fig_vibration1}(a) and  Fig.~\ref{fig_vibration2}(a), the acoustic field takes place in the whole channel. Far from the walls, fluid particles oscillate with fixed amplitude $A$ and orientation ruled by that of the nearest wall. As previously stated, while for $\lambda \gg (w, l)$ no streaming force can develop within the microchannel, the presence of a tip induces a sharp spatial gradient in the orientation of vibrations, see Fig.~\ref{fig_vibration1}(c), where the aforementioned centrifugal effect clearly appears in the vicinity of the tip. This effect induces a locally strong streaming jet right from the tip, as shown in Fig.~\ref{fig_vibration2}(b). 

Careful high-speed Particle Image Velocimetry (PIV) measurements of the acoustic flow reveal that oscillations close to the tip are stronger than elsewhere in the channel, roughly by a factor of two to three. The exact value of this factor is found to depend on both $\alpha$ and $v_{a}$, and presumably on the height $h$. 

Fig.~\ref{fig_vibration1}-(b) and (c) respectively show the amplitude of the acoustic velocity $v_a$ and the vibration orientation, quantified by the angle $\alpha_{vb}$, obtained from both experiments and simulations. Approaching $x$=0, the amplitude $v_a$ sharply increases from its value far from the tip (38.5 mm/s), to reach its maximum value at $x$ = 0 (here roughly 120 mm/s) and then sharply decreases back to its value at infinity, see Fig.~\ref{fig_vibration1}(b). The values of the velocity amplitude $v_a$ and angle $\alpha_{vb}$ are respectively symmetrical and antisymmetrical about $x$ = 0, along the vertical direction from the tip. For both quantities, the influence of the sharp structure is significant mainly within the region from $x$=-0.2 mm to 0.2 mm, hence comparable to the height of the structure $h$=0.18 mm. 

As shown in Fig.~\ref{fig_vibration1}(c), the orientation angle $\alpha_{vb}$ of $\textbf{v}_a$ varies along the $x$ direction. The evolution of $\alpha_{vb} (x)$ depends much on the distance from the tip $y_{\delta}$. If $y_{\delta}$ = 0.01 mm (line $[y_{3}-y_{3}^{\prime}]$), hence roughly equal to $\delta$, $\alpha_{vb}$ increases from 0 far enough from the tip, up to roughly $32^{\circ}$. Then it sharply decreases down to its corresponding negative value, roughly -$32^{\circ}$,  continuously and slowly increases back to zero far away from the tip.
This profile is in very good agreement with our previous measurements obtained from high-speed imaging \cite{Zhang2019} and extracted at the same distance $y_{\delta}$ from the tip. In Fig.~\ref{fig_vibration1}(c), we also plot $\alpha_{vb}(x)$ along the line $[y_{2}-y_{2}^{\prime}]$, which corresponds to $y_{\delta}$ = 0, hence intersecting the edge right at the tip. The overall profile of $\alpha_{vb}(x)$ resembles the previous one, except near the tip where the maximal and minimal values have larger absolute values, around $40^{\circ}$ and -$40^{\circ}$ respectively. Finally, the values extracted from a line $[y_{1}-y_{1}^{\prime}]$ lower than the tip, show the same  trend for $\alpha_{vb}(x)$, with maximal and minimal values very close to that of the wall, i.e. ($\pi/2 - \alpha/2$) and ($\alpha/2 - \pi/2$).

At this stage and at a qualitative level, we can conclude that the value of $\alpha$ sets the amplitude of the jump in the orientation of $\textbf{v}_a$, whereas the value of $r_c$ sets the sharpness of the spatial variation of orientation. Both of them are play crucial influence on the magnitude of acoustic perturbation into the fluid.  

\subsubsection{Streaming velocity}

Fig. \ref{fig_vibration2}(b) shows the steady streaming velocity and corresponding streamlines. We observe perfectly symmetrical streaming vortices in the vicinity of the sharp tip. This clearly shows how focused the driving streaming force is, in particular in the vicinity of the sharp tip, and confirms previous findings \cite{Ovchinnikov2014}. Thereafter, we denote the maximal value, evaluated within the whole streaming flow, as $v_{sm}$. In sharp-edge streaming, the velocity is found to be maximal along the y axis, hence at $x$=0, and directed toward the y direction. We shall see that this is no longer the case when $r_c$ is large enough with respect to $\delta$.

\begin{figure}
  \centering 
  \includegraphics[width=\linewidth]{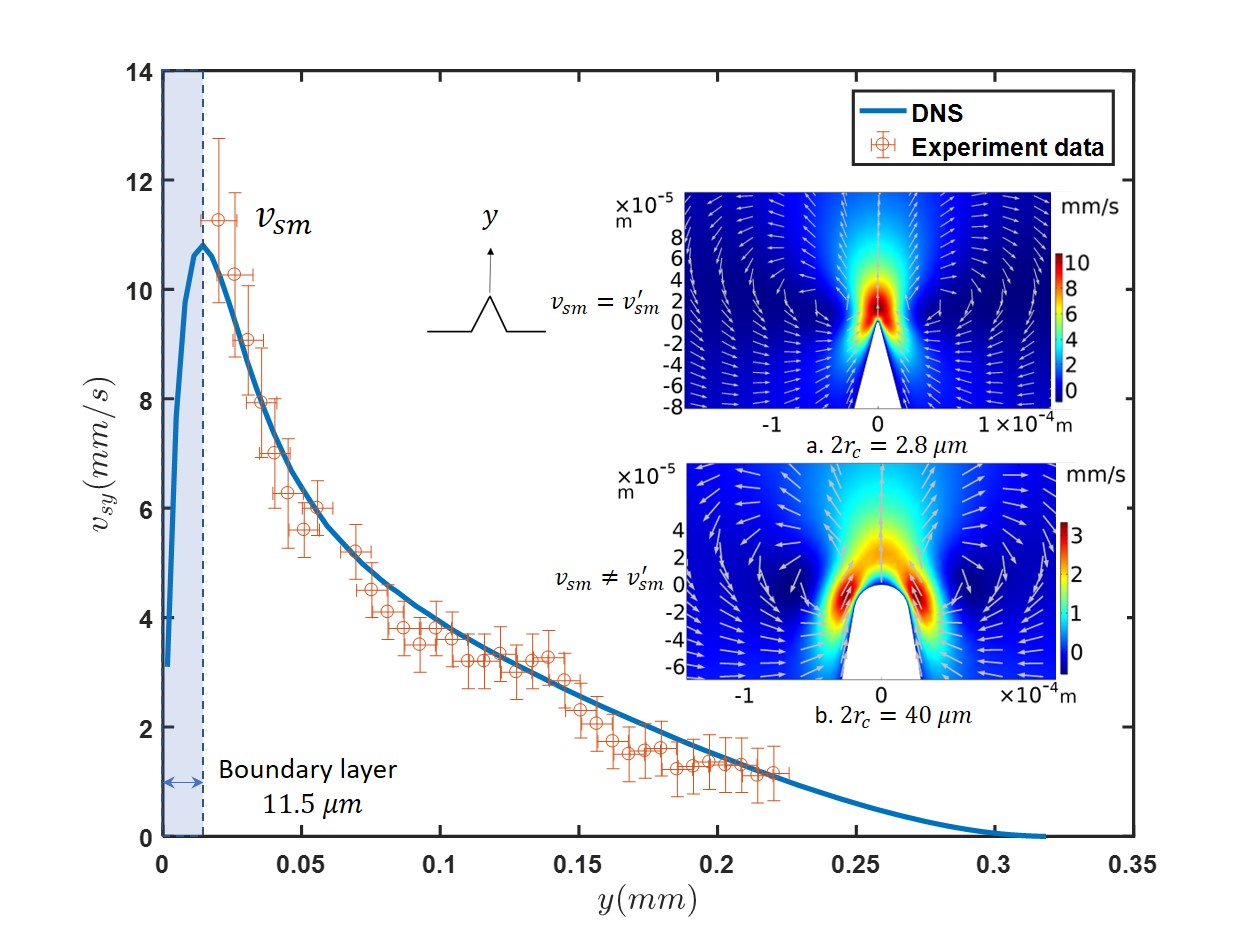}
  \caption{Streaming velocity along the y-direction: comparison between experiments and simulations. Conditions are: $\alpha$ = 30$^{\circ}$, $2r_c$=2.8 $\mu$m, $v_a$  = 37.8 mm/s. Due to the finite size of PIV particles, the flow could not be solved within the boundary layer with a thickness of 11.5 $\mu$m. The two inserted maps show the magnitude of the streaming velocity field (reddish color standing for larger velocity) and its direction (arrows) in respectively two situations: sharp edge situation (upper map) where the maximal velocity $v_{sm}$ is located on the y axis and round edge one (lower map) where $v_{sm}$ is located besides the y axis and hence different from $v_{sm}^{'}$.}
  \label{Fig_streamingYdirection}
\end{figure}

Fig. \ref{Fig_streamingYdirection} shows the streaming velocity $v_{sy} (x=0,y)$ along the $y$ direction, with the frame origin ($x$ = 0 and $y$ = 0) taken at the tip. For a reason of symmetry, $v_{sy} (x=0,y)$
is oriented along $y$ so that only the $y$ component of $v_s$ is plotted. Results from DNS are in very good agreement with experiments extracted from our previous study \cite{Zhang2019}. In addition, the numerical study further allows to access velocity within the thin VBL, which was hardly possible in experiments, due to limitations of the visualization technique. Within the VBL range $y \le \delta$, the streaming velocity sharply increases with $y$ to its maximum value $v_{sm}$ obtained near $y \simeq \delta$. Beyond this point, the streaming velocity decreases along the $y$ direction and vanishes to zero at a distance from the tip roughly equal to $w-h$, here $\simeq$ 0.3 mm.

We also define $v_{sm}^{'}$ as the maximal streaming velocity determined only on the y axis.
Let us here point out that for most situations investigated in this study, namely the situation of sharp edge where $2r_{c} < \delta$,  $v_{sm}$ is found to be along the $y$ axis (at $x$=0 and $y \simeq \delta$ like in Fig.~\ref{Fig_streamingYdirection}), and then $v_{sm}=v_{sm}^{'}$. However, when $r_{c}$ is significantly larger than $\delta$ (by a factor that remains to be determined, which quantifies the crossover between sharp-edge and classical Rayleigh streaming), the maximal velocity is found out of the $y$ axis, typically in the periphery of the two eddies of the VBL, making $v_{sm}$ different from $v_{sm}^{'}$. This is illustrated by the two insets of Fig.~\ref{Fig_streamingYdirection}. In the latter situation, these two values shall be treated separately.

\subsubsection{DNS versus PT}
Based on the above streaming analyses, we extract $v_{sm}$ as a relevant quantity to characterise the streaming velocity field, under the combination of different operating parameters. Other quantities like the size of streaming vortices and the area influenced by the streaming flow are directly related with $v_{sm}$ \cite{Zhang2019}. 
In order to better quantify the situations where $r_c > \delta$, and in particular to understand and quantify the crossover between sharp-edge and smooth-edge configurations, we also systematically extract $v_{sm}^{'}$, hence restricting the area to the $y$ axis.

\begin{figure}
    \centering
    \includegraphics[width=\linewidth]{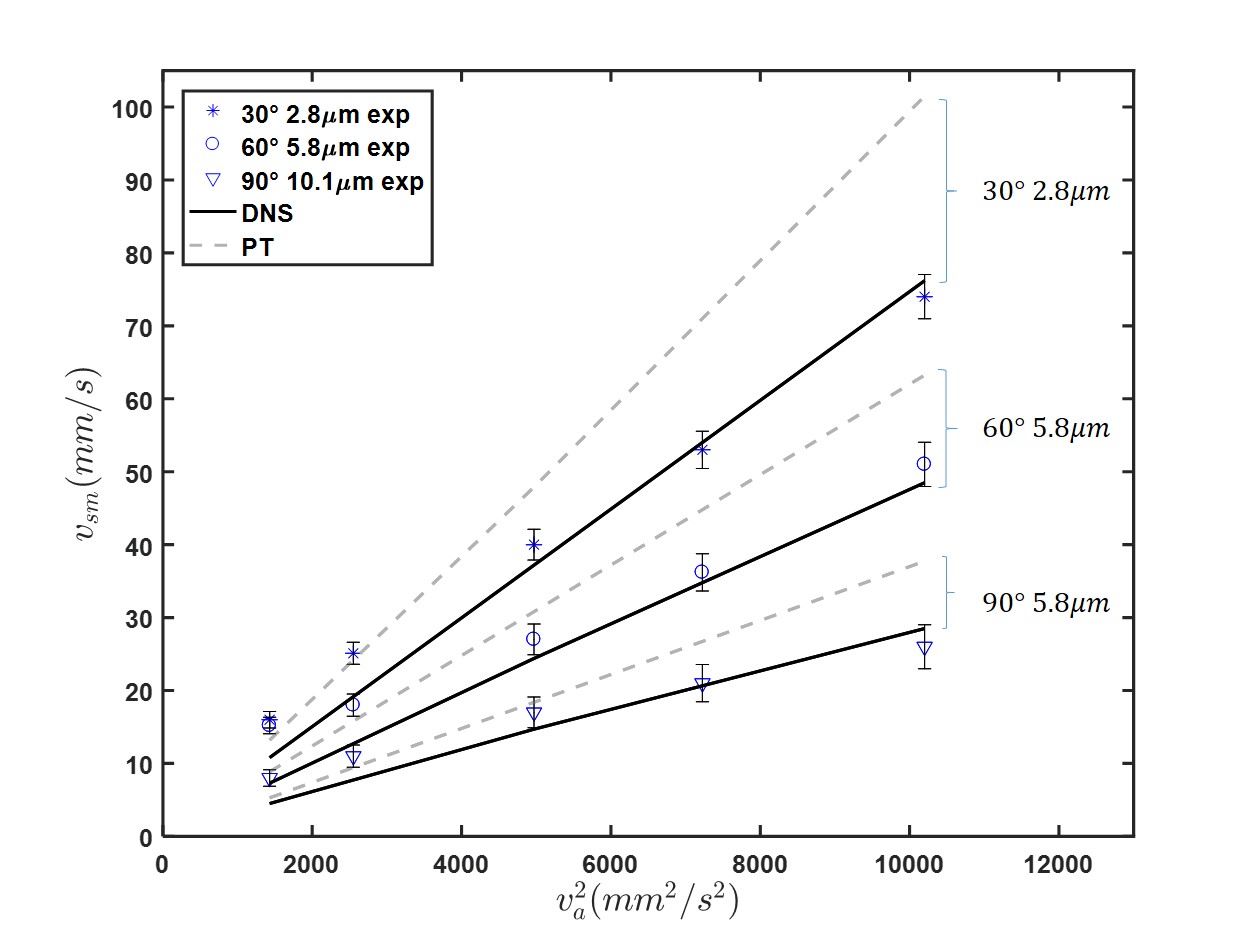}
    \caption{Maximum streaming velocity versus acoustic forcing amplitude $v_a^2$, for different sets of values for angle $\alpha$ and radius of curvature $r_c$. Results are extracted from experiments (symbols), PT simulation (dashed lines) and DNS (plain lines).}
    \label{Fig_MaxVs}
\end{figure}

We first quantitatively investigate the influence of the forcing amplitude on $v_{sm}$. Figure \ref{Fig_MaxVs} shows a quadratic dependence between $v_{sm}$ and the acoustic velocity amplitude $v_a$. Experimental results with water are taken from \cite{Zhang2019} and from three sets of values of $\alpha$ and $r_c$. Results from DNS and PT simulation are shown respectively as plain and dashed lines for the three sets of parameters. At low enough acoustic amplitude, both the PT and DNS simulations give satisfactory agreement with experiments. 

However, at larger acoustic velocity, results of DNS are in better agreement with experiments than those from PT. The latter tends to over-estimate the streaming velocity by roughly 20\% under strong acoustic vibration. 

The above results suggest that DNS provides a better prediction of the streaming velocity around the tip and it can be considered as a reliable method to predict the streaming flows generated by sharp structures.

\subsection{Quantitative results}
\subsubsection{Vorticity maps}
\begin{figure}
    \centering
    \includegraphics[width=\linewidth]{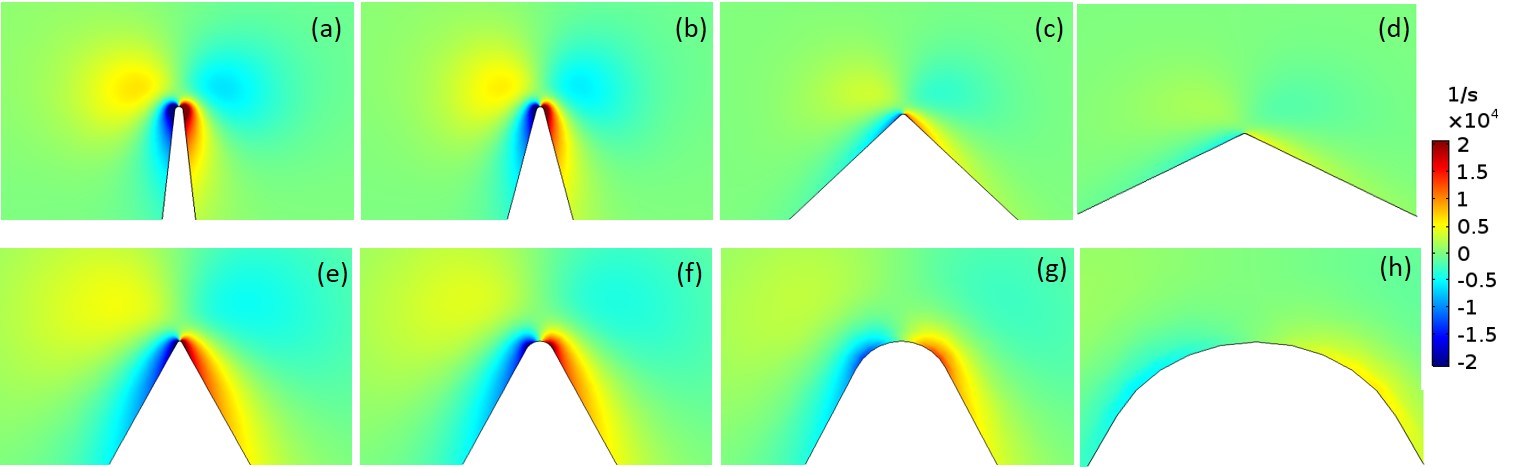}
    \caption{Vorticity maps of the streaming flow near the tip under different geometrical conditions. Red color (positive vorticity) and blue color (negative vorticity) respectively correspond to flows in anticlockwise and clockwise directions. Vibration amplitude $v_a =101.7 mm/s$. For all maps, $f$ = 2500 Hz and liquid is water, so that $\delta \simeq$ 11.3 $\mu$m. Figs.(a)-(d) have the same curvature diameter $2r_c=2.8 \mu m $ but different tip angles $\alpha$, (a) $\alpha= 12^{\circ}$; (b) $\alpha= 30^{\circ}$; (c) $\alpha= 90^{\circ}$; (d) $\alpha= 120^{\circ}$. Figs.(e)-(h) have the same tip angle ($\alpha=$ $60^{\circ}$) but different curvature diameters (e)  $2r_c$ = 1.0 $\mu$m; (f) $2r_c$ = 6 $\mu$m; (g) $2r_c$ = 20 $\mu$m; (h) $2r_c$ = 50 $\mu$m.}
    \label{Fig_vorticity}
\end{figure}

Figure \ref{Fig_vorticity} show vorticity maps of the streaming flow, calculated by DNS with different tip angle $\alpha$ and curvature diameter $2r_c$. The acoustic forcing amplitude is taken relatively strong, at $v_a=101.7 mm/s$, corresponding to the right uttermost points in Figure \ref{Fig_MaxVs}. 
It reveals that intense vorticity is localised very near to the tip, within the VBL, and takes values of opposite signs in the regions to the left and right of the tip. The inner vortices in turn induce outer vortices of opposite sign and of larger size, further away from the tip (see Subfigs (a) and (b)). These outer vortices correspond to the system of streamlines shown in Fig.~\ref{fig_vibration2}-(b). For all cases, the extrema of vorticity roughly remains at the same locations: very close to the tip in the two sides.

Subfigures (a)-(d) in Fig.~\ref{Fig_vorticity} illustrate the comparative effect at different angles (ranging from acute ($\alpha=12^{\circ}$) to obtuse ($\alpha=120^{\circ}$) edges, on the vorticity while keeping $r_c$ constant. More intense vorticity appears for sharper structures (subfigs (a-b)) while its magnitude decreases as $\alpha$ increases (subfigs (c-d)).
In turn, subfigures (e)-(h) correspond to different values of $r_c$, while keeping $\alpha$ constant. Two distinct behavior emerge: when $2r_c > \delta$, increasing $r_c$ leads to more spread and weaker vorticity, while when $2r_c < \delta$ the vorticity does not vary significantly with $r_c$.

Based on these results, we can conclude that the curvature diameter and tip angle have qualitatively different influences on the vorticity, hence on the streaming. A smaller and sharper structure provides stronger streaming force and flow. 

\subsubsection{Streaming velocity magnitude per acoustic power}

We now aim to define a simple fitting parameter to quantify the efficiency of the response of streaming flow in regards to the prescribed vibration.

The discussions in Ovchinnikov \textit{et al.} \cite{Ovchinnikov2014} describe typical streaming velocity in cylindrical coordinate $(r,\phi)$ as:

\begin{equation}
    v_s(r)=\frac{v_a^2}{\nu}\frac{\delta^{2n-1}}{a^{2n-2}}H_{\alpha}(\frac{r}{\delta})
    \label{ovchinnikov_vs}
\end{equation}

\noindent where $n$ is a coefficient that depends on $\alpha$, $n = \frac{\pi}{2 \pi - \alpha}$; $a$ is a length scale close to that of the sharp-edge height $h$. The function $H_{\alpha}(\frac{r}{\delta})$ contains the radial profile of the streaming flow. Quantitatively, we mainly focus on the characteristic (and maximal) value of $v_s(r)$ at $r=\delta$ and $\phi$ = 0, so what follows we shall just consider the constant prefactor $\frac{1}{\nu}\frac{\delta^{2n-1}}{a^{2n-2}}$ that relates $v_s$ to $v_a^2$. Let us note that this equation, supposedly valid in the range $r_c < \delta$, does not contain any dependence on $r_c$. 

The results presented in Figure \ref{Fig_MaxVs} confirm that for a given combination of geometry, acoustic frequency and liquid viscosity - and actually most of the experimentally relevant conditions, $v_\text{sm}$ varies quadratically with the amplitude of vibration velocity $v_a$. Therefore, we define the fitting parameter $\theta = \frac{\Delta v_\text{sm}}{\Delta (v_a^2)}$ as a measurement of the efficiency of the momentum conversion from acoustic to streaming flows. We shall consider this parameter in the following discussions in order to quantify the influence of the different varying parameters namely $\alpha$, $r_c$ and $\nu$. Similarly, we define $\theta^{'} = \frac{\Delta v_\text{sm}^{'}}{\Delta (v_a^2)}$.

\subsubsection{Influence of tip angle}
\begin{figure}
    \centering
    \includegraphics[width=\linewidth]{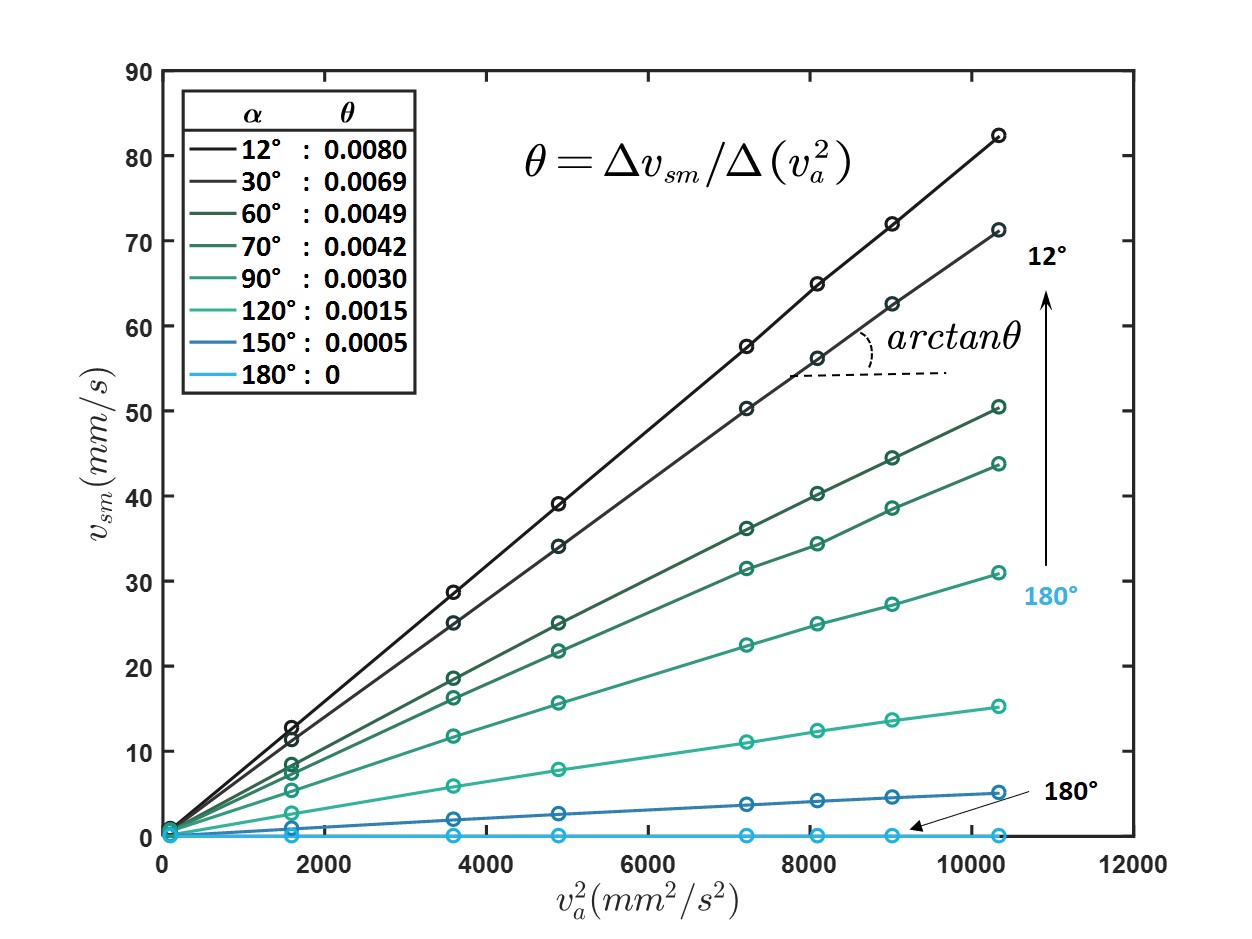}
    \caption{Maximum velocity versus square of the vibration amplitude $v_a^2$, with different tip angles. The coefficient $\theta$ is extracted from a linear fit, which holds very well within the whole range of $v_a$. Other conditions are: $2r_c$ = 2.8 $\mu$m, $f$ = 2500 Hz.}
    \label{Fig_tipangle}
\end{figure}

In this first series of results, we quantify the strength of the streaming flow for different values of the angle $\alpha$, from 12$^{\circ}$ to 180$^{\circ}$, keeping all other quantities constant. Figure \ref{Fig_tipangle} shows $v_\text{sm}$ versus $v_a^2$ for different values of $\alpha$. In particular, as illustrated in vorticity maps of the Figures \ref{Fig_vorticity}-(a-d), $r_c$ can be kept constant for different $\alpha$, except of course for $\alpha$ = 180$^{\circ}$ that corresponds to case of a flat, straight wall. As previously stated, the more acute the angle, the stronger the streaming flow for a given $v_a$. Besides, a flat wall with $\alpha$ = 180$^{\circ}$ does not generate any streaming flow even for strong $v_a$.

\begin{figure}
    \centering
    \includegraphics[width=\linewidth]{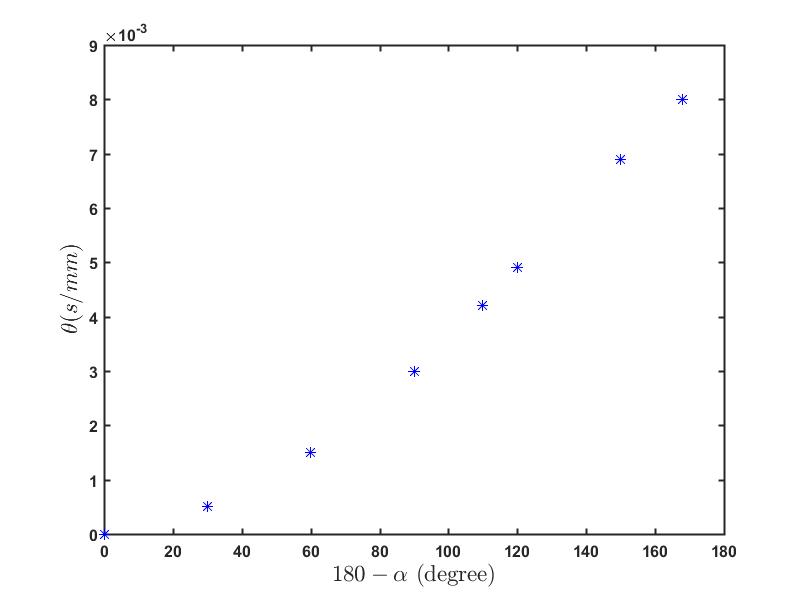}
    \caption{Coefficient $\theta$ versus the supplementary of the tip angle $180 - \alpha$. Other conditions are: $2r_c$=2.8$\mu$m (except for $\alpha$ = 180$^{\circ}$ where $r_c$ is infinite), $f$=2500 Hz.}
    \label{Fig_tipanglevstheta}
\end{figure}

Since the vast majority of cases exhibited a robust quadratic dependence between $v_\text{sm}$ and $v_a$, we extracted $\theta$ for each value of $\alpha$. The result is shown in Figure \ref{Fig_tipanglevstheta}, where $\theta$ is plotted versus $(180^{\circ}- \alpha)$. 

Under fixed values for other parameters, here, $2r_c$ = 2.8 $\mu$m, $f$ = 2500 Hz, $\theta$ achieves its highest value with the sharpest angle, $\alpha = 12^{\circ}$. The maximal efficiency of the momentum conversion is slightly below 10$^{-2}$ s/mm. When $\alpha$ increases to 90$^{\circ}$, $\theta$ drops to roughly 3$\times$10$^{-3}$ s/mm, and it vanishes to zero when $\alpha$ approaches 180$^{\circ}$. The dependence of $\theta$ with $\pi - \alpha$ is then strongly non-linear.

\subsubsection{Influence of tip curvature}
\begin{figure}
    \centering
    \includegraphics[width=\linewidth]{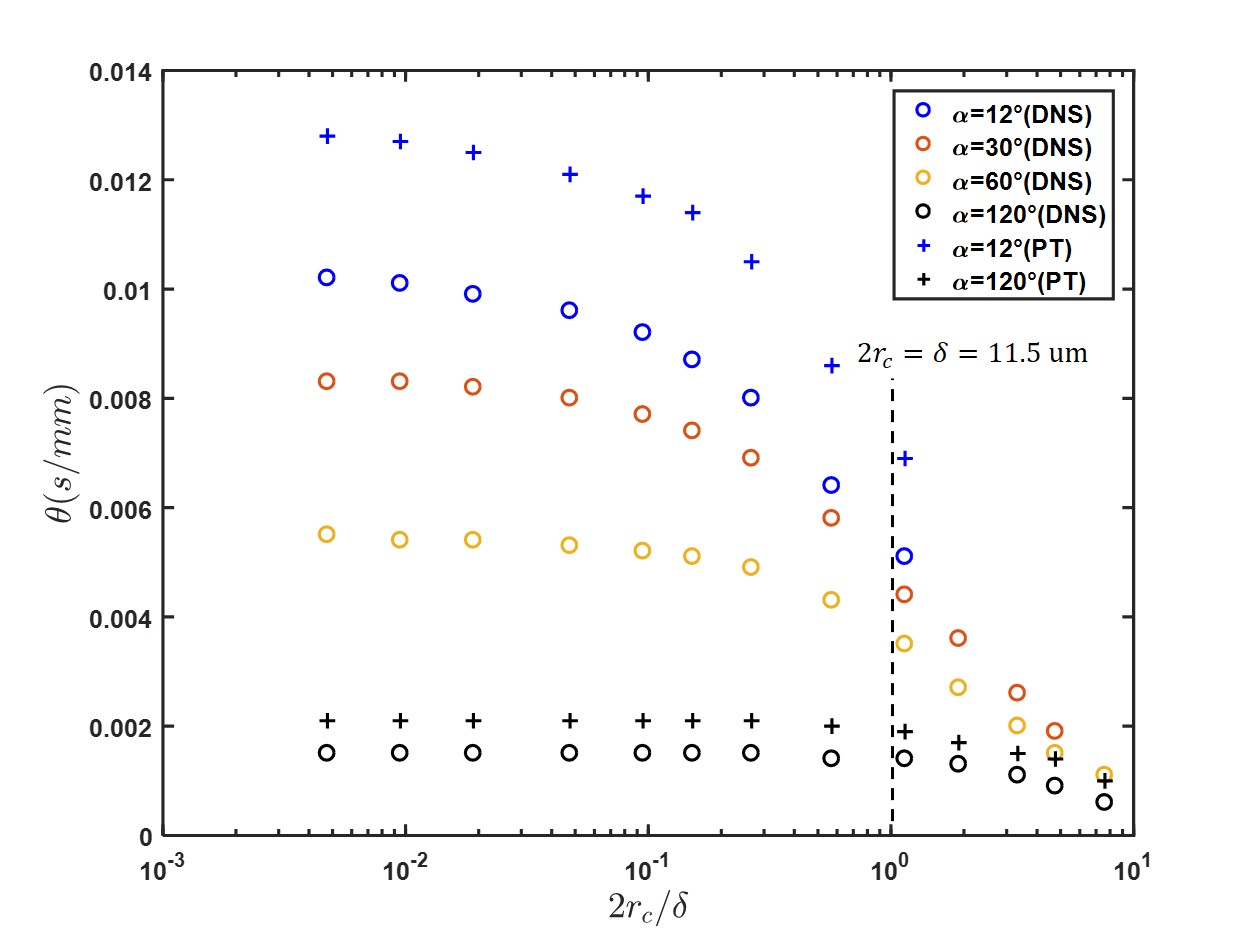}
    \caption{Coefficient $\theta$, based on the maximal value of streaming velocity, versus ratio between curvature diameter $2r_c$ and boundary layer thickness $\delta$, for four different tip angles $\alpha$. DNS results should be considered as reliable and PT simulation appears to over-estimate the result according to the two extreme cases ($\alpha$ = 12$^{\circ}$ and $\alpha$ = 120$^{\circ}$).}
    \label{Fig_rtodelta}
\end{figure}

We now investigate the influence of $r_c$ on $\theta$, for a series of four $\alpha$ values from 12$^{\circ}$ to 120$^{\circ}$. Simulations were carried out under the same liquid viscosity (water, $\nu$=10$^{-6}$ m$^2$/s) and frequency $f$ = 2500 Hz, so that $\delta$ was kept constant at 11.5 $\mu$m and only $r_c$ was varied. Figure \ref{Fig_rtodelta} shows the dependence of $\theta$ versus $2 r_c/\delta$. 

These results reveal a decrease of $\theta$ with $r_c$, and this decrease becomes more significant within the range $2r_c > \delta$, see Figure \ref{Fig_rtodelta}. Hence, the conversion of acoustic power into streaming flow is less efficient when $r_c$ becomes large.
Let us note that in Figure \ref{Fig_rtodelta}, we also put results from simulations using the PT method for the two extreme values of $\alpha$, again in the aim to illustrate the gap between both methods. It confirms that PT systematically over-estimates the magnitude of the streaming flow, by a factor of roughly 1.2. 

This constitutes a quantitative confirmation of what was suggested in the vorticity maps of Figures \ref{Fig_vorticity}(a-d). Also, the influence of $r_c$ is more pronounced when the tip angle is more acute. 

Once $2r_c$ is increased larger than $\delta$, $\theta$ significantly decreases, which is common for all tip angles (Fig.\ref{Fig_rtodelta}). This is in accordance with the spreading and weakening contour observed in Figure \ref{Fig_vorticity}-(g-h). When the tip is no longer sharp, the magnitude of AS weakens as we should retrieve the classical Rayleigh-Schlichting streaming.

\subsubsection{Influence of viscosity}
One of the remarkable and non-intuitive features of Rayleigh-Schlichting streaming is its independence on viscosity, providing that the typical size of the container is much larger than the thickness of the VBL, $\delta$ \cite{Westervelt1953,Nyborg1958}. This classical result, which expresses that streaming is both spawned and hindered by viscosity, can be retrieved by simple scaling arguments \cite{Boluriaan2003a,Costalonga2015}, though it is no longer true in confined geometries \cite{Costalonga2015}. Here in the case of sharp-edge streaming, we show that, despite $\delta$ can remain small compared to the channel size, viscosity has a strong influence on the sharp-edge induced streaming.

\begin{figure}
    \centering
    \includegraphics[width=\linewidth]{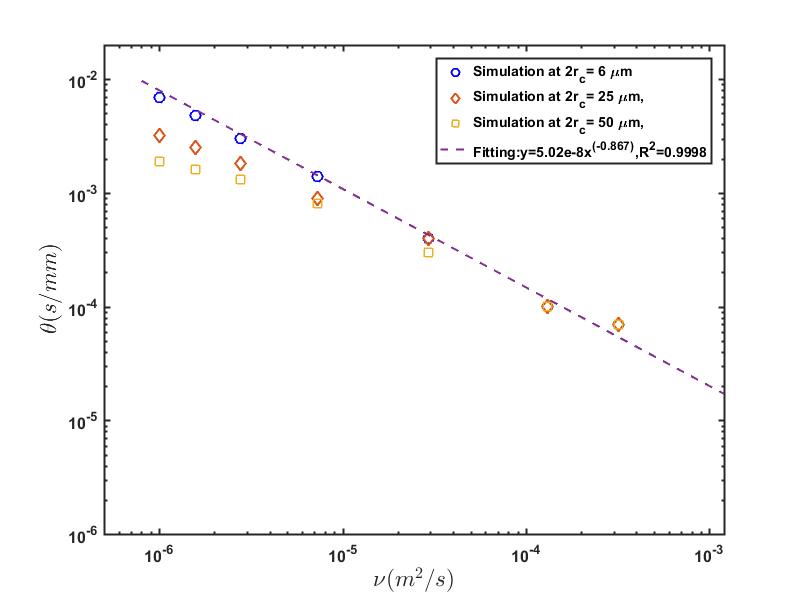}
    \caption{Coefficient $\theta$ versus kinematic viscosity $\nu$. Other parameters are: $\alpha$ = 30$^{\circ}$, $f$ = 2500 Hz. The fitting power-law curve is based on the results for $2r_{c}=6 \mu m$. For $\nu >$ 5$\times$10$^{-5}$ m$^2$/s, the data points are all at the same value, showing that $\theta$ is almost independent on $r_c$.}
    \label{Fig_viscosity}
\end{figure}

Figure \ref{Fig_viscosity} shows a strong decrease of $\theta$ with kinematic viscosity $\nu$ in Log-Log axes. We span a large range of values for $\nu$, from that of water (10$^{-6}$ m$^2$/s) to a 1000-times more viscous liquid, with a corresponding $\delta \simeq$ 357 $\mu$m, which in practice would correspond for instance to pure glycerin. By increasing $r_c$ from 6 $\mu$m to 50 $\mu$m, the evolution of $\theta$ shows a crossover from sharp-edge streaming to classical Rayleigh-Schlichting streaming. In particular for the sharp edge with $2r_c$ = 6 $\mu$m, together with $f$ = 2500 Hz and constant $\alpha$ = 30$^{\circ}$, we remain in a sharp-edge streaming situation since $2r_c < \delta$. The decrease can be well fitted by a power-law, with an exponent of -0.867 giving the best fit, see Fig.~\ref{Fig_viscosity}. The cases investigated with $2r_c$= 25 and 50 $\mu$m reveal that the decrease of $\theta$ with viscosity is much less pronounced for higher values of $2r_c /\delta$, hence in the lower viscosity range. Therefore if $2r_c  > \delta$, it turns out that the dependence of $\theta$ on viscosity is not captured by a power law. Let us note that, while we ran our simulations up to $\nu$=10$^{-3}$ m$^{2}$/s, the relationship between $v_{sm}$ and $v_{a}^{2}$ is no longer purely linear within this high-viscosity range. Therefore, the value of $\theta$ could not be extracted for the highest values of $\nu$.

Conversely, the value of $\theta$ is independent on $r_c$ in the high viscosity range, i.e. when $2r_c /\delta <$ 1: this is a trademark of sharp-edge streaming.  

\begin{figure}
    \centering
    \includegraphics[width=\linewidth]{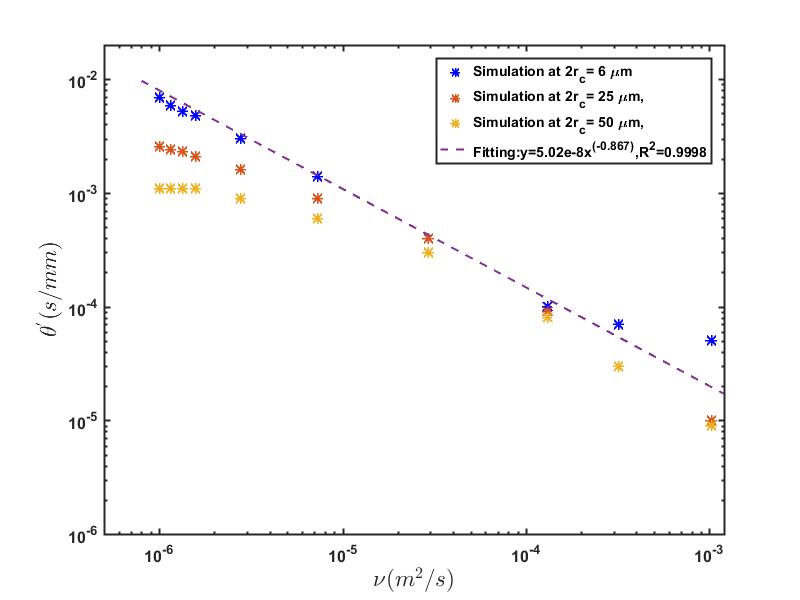}
    \caption{Coefficient $\theta^{'}$ versus kinematic viscosity $\nu$. Other parameters are: $\alpha$ = 30$^{\circ}$, $f$ = 2500 Hz. The fitting power-law curve is based on the simulation when $2r_{c}$=6 $\mu$m.}
    \label{Fig_viscosity_vymax}
\end{figure}

Equation (\ref{ovchinnikov_vs}) taken from Ovchinnikov \textit{et al.}'s study \cite{Ovchinnikov2014} predicts a decrease of $\theta$ with $\nu$ via a power-law of negative exponent, as $\theta \sim \nu^{\left(n-\frac{3}{2}\right)}$. For the chosen angle $\theta$ = 30$^{\circ}$, $n \simeq$ 0.54, yielding an exponent $\left(n-\frac{3}{2}\right)$ of -0.96, close to but different from the value of -0.867 found empirically.

The evolution of $\theta^{'}$ with $\nu$ also shows a global decrease, see Fig.~\ref{Fig_viscosity_vymax}. But the main difference with $\theta$, is that within the range where $r_c / \delta$ is large enough, the value of $\theta^{'}$ is independent on $\nu$. It is clearly evidenced for $r_c$ = 50 $\mu$m in the low viscosity range. The distinction between $\theta$ and $\theta^{'}$ is mostly significant where $r_c / \delta >$ 1, which corresponds to situations depicted in the vorticity maps of Figs.~\ref{Fig_vorticity}-(g,h). In these situations, the maximum of streaming velocity is not localized along the $y$ axis. Still, the behavior of $v_{sm}^{'}$ follows a quadratic increase with $v_a$, so that $\theta^{'}$ remains well defined. More surprisingly, beyond $\nu$=10$^{-4}$ m$^2$/s, the decrease of $\theta^{'}$ with $\nu$ deviates from a power law. Also, $\theta^{'}$ remains dependent on $r_c$ in the whole range of $\nu$ investigated.

\section{Discussions and conclusion}
Let us now recall and summarize the main results. Motivated by experimental results on the generation of intense acoustic streaming near sharp edges \cite{Huang2013a,Huang2014,Ovchinnikov2014,Nama2014,Nama2016a,Zhang2019}, the results of our DNS simulations allow a characterization of the streaming flow both outside and inside the VBL, of typical thickness $\delta$ = 11.5 $\mu$m for water at $f$ = 2500 Hz. This constitutes a significant step forward with respect to the state of the art, since  it is experimentally hard to access the flow details within the VBL \cite{Zhang2019} and few studies employs the DNS method, which is shown to be of higher precision than classical PT. In particular, providing $2r_c$ is smaller than $\delta$ (which is the case of sharp-edge streaming), the maximum of streaming velocity is found near the apex of the sharp tip, at a distance of roughly $y=\delta$, inducing regions of strong and concentrated vorticity aside and within the VBL, as well as larger outer vortices that can ensure efficient mixing across the whole channel \cite{Huang2013a,Nama2014,Nama2016a,Zhang2019}.

Furthermore, we gained better understanding of the first-order acoustic velocity field. It turns out that both the orientation of the oscillation and its amplitude are \textit{tuned} by the sharp structure to give a strongly localized perturbation to the fluid. Namely, the norm of ${\bf{v}}_a (x,y)$ is maximal near the tip, precisely around the location of the maximum of streaming velocity $(x=0, y=\delta)$. Surprisingly, this amplitude $v_a$ is found twice to three times larger than that far away from the tip. Let us note that this confirms recent experiments \cite{Zhang2019}. 
This effect, which significantly contributes to the streaming efficiency, depends on the sharp edge structure. Though, it remains to be explained and quantified in more details.

Our study also focuses on the influence of the tip sharpness, and reveals that the two parameters $r_c$ and $\alpha$ are crucial for the generation of acoustic streaming. While their respective influences were difficult to dismantle in experiments, our numerical results provided a better understanding. Since the acoustic flow direction (angle $\alpha_{vb}$) follows that of the walls, the sudden change of oscillation orientation beside the tip leads to centrifugal effect. Along a typical length as short as $2r_c$, the orientation jumps from $\alpha_{vb} = \pi/2 - \alpha/2$ for $x \ll r_c$ to $\alpha_{vb} = \alpha/2 - \pi/2$ for $x \gg r_c$, hence a total turn of $\Delta\alpha_{vb} = \pi - \alpha$. This \textit{gradient} generates strong values for the effective streaming force $F_s$. 

Let us briefly comment on Eq.~(\ref{timeaveragedns}). Due to relatively local strong values for $\textbf{v}_s$, i.e. comparable in magnitude to $\textbf{v}_a$, the non-linear term $(\textbf{v}_s \cdot \nabla)\textbf{v}_s$ should in principle be significant. The physical meaning of this term can be viewed as the self-advection of the streaming flow, which in practice leads to vortex elongation in Rayleigh-Schlichting streaming \cite{bahrani2020vortex}. However, this is somewhat contradictory with the robust quadratic relationship found between $v_{sm}$ and $v_a$, regarding Eq.~(\ref{timeaveragedns}). To explain this apparent contradiction, we retain two possible hypotheses:

i) although the magnitude of $\textbf{v}_s$ can locally be relatively large, the term $(\textbf{v}_s \cdot \nabla)\textbf{v}_s$ could be negligible, especially in the region around the maximum $v_{sm}$.

ii) the term $(\textbf{v}_s \cdot \nabla)\textbf{v}_s$ could be irrotational, so that it can be exactly compensated by the pressure gradient term $\nabla p_s$.

To test these two assumptions, we plotted the maps of the norms of both quantities $(\textbf{v}_s \cdot \nabla)\textbf{v}_s$ and $\nu \nabla^2 \textbf{v}_s$. The results are shown in Fig.~\ref{figtermscompare}, for a typical value of $v_a$ in the intermediate range. It turns out that the first assumption is the right one, as it shows that the magnitude of $\left\lVert \nu \nabla^{2} \textbf{v}_{s}\right\rVert$ overcomes that of $\left\lVert (\textbf{v}_{s}\cdot \nabla)\textbf{v}_{s}\right\rVert$ by a factor of roughly 70. Therefore, the non linear term $(\textbf{v}_s \cdot \nabla)\textbf{v}_s$ can be considered as negligible in Eq.~(\ref{timeaveragedns}), which explains the extension of the quadratic behavior between $v_{sm}$ and $v_a$ in sharp-edge streaming. Furthermore, it underlines that the differences between PT and DNS simulation results, and the fact that DNS matches better experiments of sharp-edge streaming should be explained by the importance of the other non-linear terms $(\textbf{v}_a \cdot \nabla)\textbf{v}_s$ and $(\textbf{v}_s \cdot \nabla)\textbf{v}_a$ in Eq.~(\ref{nsosc}). 

In the seek for optimal operating conditions of sharp edge AS, the efficiency of conversion from acoustic vibrations to streaming flow is quantified by $\theta$. In particular, while the fabrication of sharp tips requires in practice careful and expensive techniques, especially for $r_c$ as small as a few microns, Fig. \ref{Fig_rtodelta} shows that the streaming flow does not gain much in strength when $r_c$ is lowered below $\delta / 2$. The precise identification of the influence of $r_c$ and $\alpha$ was made possible thanks to the DNS simulations.

The role of viscosity was also investigated. The power-law decrease of $\theta$ with $\nu$, predicted by Ovchinnikov \textit{et al.} \cite{Ovchinnikov2014} was confirmed by our simulations, although the exponent was found weaker than the predicted one. We also confirmed the independence of $\theta$ on $r_c$ in the range $2r_c < \delta$, and we investigated the crossover between the sharp edge AS and classical Rayleigh streaming regimes by tuning the value of $2r_c/\delta$. In particular, we recover the independence of $\theta$ on $\nu$ if $2r_c/\delta \gg 1$. 

Also, our simulations showed that the quadratic relationship $v_{sm} \sim v_a^2$ fails for a high enough viscosity. This has to be considered as a limitation of the geometry since with higher range of $v_a$, the size of the outer vortices is comparable to that of the channel width $w$. The streaming is thus limited by the size of the microfluidic channel.

Let us also suggest a quantitative criterion of efficiency in the context of (macro-)mixing under a typical imposed flow-rate $Q$ through the channel. Previous experiments quantifying both the maximal streaming velocity and mixing efficiency revealed that a satisfying mixing rate could be obtained if the averaged flow velocity, here $<V> = \frac{Q}{w\cdot d}$, is comparable to the maximal streaming velocity \cite{Zhang2019}. For the width $w$ = 500 $\mu$m and depth $d$ = 50 $\mu$m used in these experiments, and a middle-range value of $Q$ = 10 $\mu$l/mn, it yields: $<V> \simeq$ 6.7 mm/s. Therefore in practice, the setpoint velocity $v_{sm} = <V>$ shall be related to specific conditions on both the geometry of the tip and liquid viscosity ruling the value of $\theta$, also taking into account the maximal $v_a$ that the transducer can generate. It worth noting though, that micro-mixing at the molecular scale also depends on the form of vortices generated by AS and one of our upcoming study addresses this issue using Iodate-Iodide reaction as a chemical probe \cite{Guo2013}. 

Concerning the 2D approximation, which is necessary for the DNS simulation due to that the 3D equivalent would be much more expensive in terms of computational cost, we observed little influence on the results. This is partially thanks to the high channel width/depth ratio that let us focus on the width-length plane other than width-length-depth volume. 

To sum up, the FEM-based DNS method gives very satisfactory agreement with experimental results and it over-performs the classical PT model. The latter does not consider the non-linear terms in the streaming force calculation and tends to over-estimate the sharp edge streaming. In this sense, our study shows that, providing the right boundary conditions are prescribed and all non-linear terms are kept in the calculation, AS streaming can be successfully studied in a quantitative way, with minimal inexpensive computing material (i.e. without computer cluster nor MPI) and a FEM commercial software.

\section{Appendix} \label{appendix}

\subsection{Perturbation Theory and its implementation}

The Perturbation Theory is generally adapted to address acoustofluidics problems in the framework of ``weak disturbance''. With limited access to Computational Fluid Dynamics (CFD), PT is a very powerful tool to reduce the N-S equation, which potentially include non-linear terms that couple the acoustic and streaming velocity fields, into a simpler one where only the non-linear term involving the acoustic velocity within the VBL remains significant. Therefore, PT provides an convenient method to bring out the physical fundamental core of the acoustic streaming problems while retaining relatively simple mathematical formulation \cite{Ovchinnikov2014,Bruus2012d,Sadhal2012,Lighthill1978,Riley1998a,Nyborg1953}.

For the present study, $v_a$, $v_s$ are governed by both Eq.~(\ref{nsosc}) and Eq.~(\ref{nssteady}), which set respectively the oscillating and steady terms in the velocity fields. The PT assumes $v_a\gg v_s$ so that the inertial terms in the Eq.~(\ref{nsosc}), $(\textbf{v}_s \cdot \nabla)\textbf{v}_a$ and $(\textbf{v}_a \cdot \nabla)\textbf{v}_s$, can be neglected. Without these terms, Eq.~(\ref{nsosc}) and Eq.~(\ref{nssteady}) can then be solved separately to obtain $v_a$ and $v_s$.

The procedure of the calculation based on PT can be proceeded by two steps: 
i) Solving the wave equation Eq.~(\ref{nsosc}) to determine the vibration velocity field in the geometry structure, with first-order time-periodic terms, and ii) Solving the streaming equation Eq.~(\ref{nssteady}), in which the force term in Eq.~(\ref{bodyforce}) can be determined by the results of the previous step. The second-order terms are steady ones, from which the streaming velocity $v_s$ is deduced.

Although Ovchinnikov \textit{et al.} \cite{Ovchinnikov2014} points out the limitation of PT with respect to DNS method, PT remains a powerful framework to analyze the underlying physics of the streaming fields near the sharp tip, especially when the vibration amplitude within the liquid is small enough so that the acoustic Reynolds number $Re_a = \frac{A \omega h}{\nu}$ remains of the order of one.

In COMSOL, basic steps to implement Perturbation Theory are:\par
\begin{enumerate}
    \item Module ``Thermoviscous Acoustics, Frequency Domain''for solving the acoustic vibration velocity field;
    \item Module ``Laminar flow'' for solving the streaming velocity field with $ \textbf{F}_s = - \frac{\rho}{2} \left<  Re[(\textbf{v}_a \cdot \nabla)\textbf{v}_{a}^*] \right>$ as the ``Volume Force'' inserted into the model;
    \item Boundary conditions: 
To solve the vibration velocity, the left and right boundaries (labelled as 1 and 6) are set with the acoustic velocity oscillating at the prescribed value of amplitude in the normal direction, and to be in phase with each other Other boundaries are set as no-slip walls.  

For the second-order streaming velocity, the left and right sides of the domain are set as inlet and outlet at given incoming velocity, here taken equal to zero. The other boundaries are set to be no-slip walls.
\end{enumerate}

\subsection{Direct Numerical Simulation implementation}

The detailed description of DNS has been given in Section \ref{theory}. Implementing DNS is COMSOL includes the following steps: 

\begin{enumerate}
    \item Module ``Laminar Flow'' for direct solving the N-S equations with periodic velocity boundary conditions; 
    \item Module ``Domain ODEs and DAEs'' for calculating the time average values of the velocity field in step 1; 
    \item Boundary conditions: 
The acoustic velocity (in form of a sinusoidal function of time) is set as the left boundary condition and the right boundary condition is set as a pressure $p_0$. Other boundaries are set as no-slip walls.
\end{enumerate}

\subsection{Mesh and grid independence study}

The mesh grid is built with triangle elements, with the maximum element size being 0.014 mm, and the minimum one being 0.0002 mm. Smooth transition is performed with a maximum element growth rate of 1.1. Close to the sharp edge, the mesh is refined by inflation layers to better account for the strong velocity gradients inside the VBL. The number of the layers is 3 and the layer stretching factor is 1.2.

The mesh independence is assessed by comparing the results from the chosen mesh with those obtained in a refined mesh, which is generated by increasing the number of cells by 30\%. Comparing the two meshes, the obtained streaming velocity value differs by less than 1\%. The current mesh is thus considered as being as satisfactory balance between both in terms of accuracy, reliability and computing time. 

\subsection{Time to reach steady streaming field and time step}

\begin{figure}
    \centering
    \includegraphics[width=\linewidth]{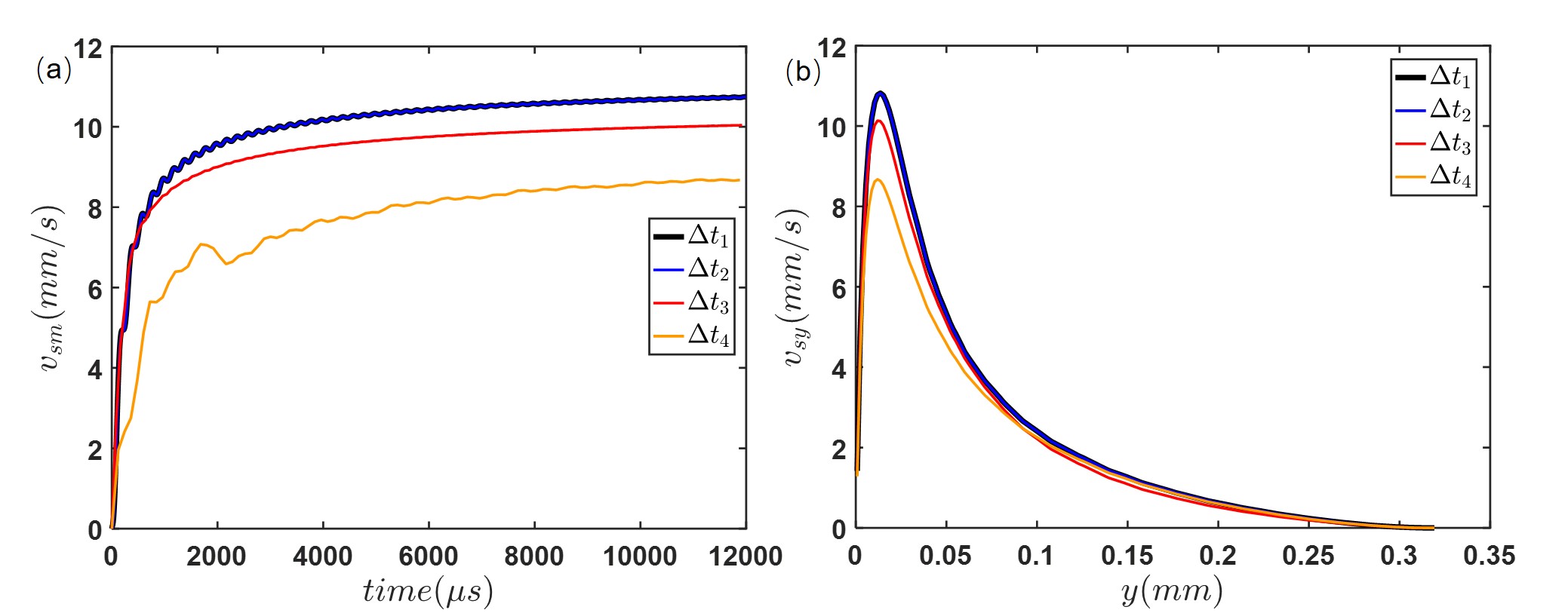}
    \caption{Variation of streaming velocity with numerical iteration time (a), and the steady y-direction streaming velocity at different time steps (b). Time steps $\Delta t_{1} =1  \mu s$, $\Delta t_{2} =8  \mu s$, $\Delta t_{3} =80  \mu s$, $\Delta t_{4} =120  \mu s$, correspond to 1/400$^{th}$, 1/50$^{th}$,1/5$^{th}$ and 1/3.33$^{rd}$ of an acoustic period. The whole duration of the simulation equals 30 acoustic periods.}
    \label{figgridtimeinde}
\end{figure}

For the PT method, the two-steps procedure belongs to a steady computation process which can be done almost instantaneously by computer. However, the streaming flow appears after a transient state, and thus needs some time to be fully developed and reach its steady state. As shown in Fig.~\ref{figgridtimeinde} (a), the streaming velocity $v_{sm}$ (the time average of the total velocity from the beginning of the simulation to a given time) grows with the number time-steps until reaching a steady state. The corresponding time duration is roughly 12 $ms$, hence 30 acoustic cycles under the acoustic frequency of 2500 Hz (period of 400 $\mu s$).

The value of the time-step is also essential to meet the CFL (Courant-Friedrichs-Lewy) stability condition. The Courant number, given by $CFL={v_a \Delta t}/{\Delta x}$, should be kept lower than 1 to guarantee the numerical iteration stable \cite{Muller2015}. As shown in Fig.\ref{figgridtimeinde}, we test four time-steps from 1 $\mu$s to 120 $\mu$s, or from 1/400$^{th}$ to 3/10$^{th}$ of an acoustic period. Only $\Delta t_{4} =120  \mu s$ gives a CFL higher than unit but $\Delta t_{3} =80  \mu s$ is not fine enough to give a satisfactory maximum streaming velocity $v_{sm}$, see Fig.~\ref{figgridtimeinde}-(a) and a reliable streaming distribution along the y direction $v_{sy} (y)$, see Fig.~\ref{figgridtimeinde}-(b). We thus choose $\Delta t_{2} =8  \mu s$ as a compromise since it gives the same results as $\Delta t_{1} =1  \mu s$ but with a shorter computing time. 

With the chosen time step of 8 $\mu$s and a total of 30 acoustic cycles, the DNS computing cost is about 25 mn per case study on an Intel i5-7500 CPU and 16G RAM.

\par

\subsection{Convective v.s. viscous terms} 
\begin{figure}
    \centering
    \includegraphics[width=\linewidth]{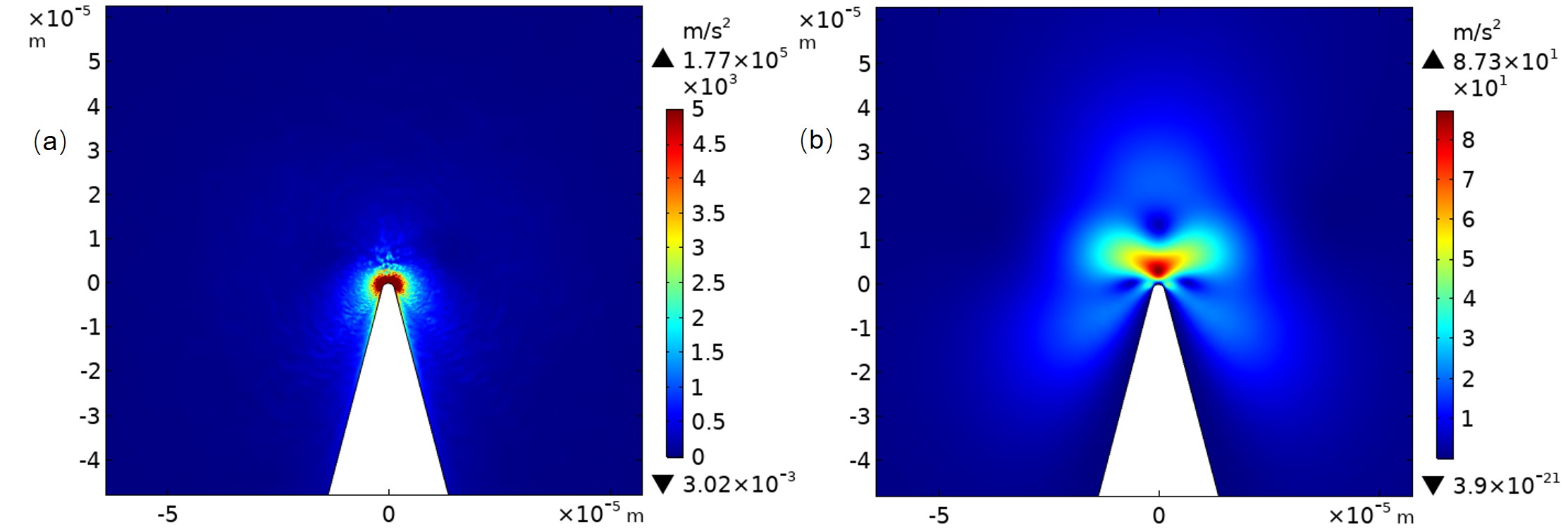}
    \caption{Intensity map of $\left\lVert \nu \nabla^{2} \textbf{v}_{s}\right\rVert$  (a) and $\left\lVert (\textbf{v}_{s}\cdot \nabla)\textbf{v}_{s}\right\rVert$ (b) at $v_a=70.5 mm/s$. The left term has a maximum of 1.77 $\times 10^5 m/s^2$, which is roughly 2000 times stronger than the right one (87.3 $m/s^2$). Close to the tip, the ratio is about 70. $(\textbf{v}_{s}\cdot \nabla)\textbf{v}_{s}$ can thus be considered as negligible in Eq.\ref{nssteady}}
    \label{figtermscompare}
\end{figure}

Equation (\ref{nssteady}) suggests that the quadratic dependence of $v_{sm}$ with $v_a$ should be right only if the term $(\textbf{v}_s \cdot \nabla)\textbf{v}_s$ is negligible compared to the other ones. Therefore, we compared the relative magnitude of $\left\lVert \nu \nabla^{2} \textbf{v}_{s}\right\rVert$ and $\left\lVert (\textbf{v}_{s}\cdot \nabla)\textbf{v}_{s}\right\rVert$, in the form of colormaps shown in Figure \ref{figtermscompare}. The chosen $v_a$ = 70.5 mm/s corresponds to a value in the median range of investigation, but this remains true even for the largest investigated $v_a$, i.e. 107 mm/s. This confirms that although $\textbf{v}_s$ can be comparable to $\textbf{v}_a$ in magnitude, the term $(\textbf{v}_s \cdot \nabla)\textbf{v}_s$ remains small compared to the others of Equation (\ref{nssteady}).

\bibliographystyle{apsrev4-1}
\bibliography{biblio_simusNS}

\begin{thebibliography}{51}%
\makeatletter
\providecommand \@ifxundefined [1]{%
 \@ifx{#1\undefined}
}%
\providecommand \@ifnum [1]{%
 \ifnum #1\expandafter \@firstoftwo
 \else \expandafter \@secondoftwo
 \fi
}%
\providecommand \@ifx [1]{%
 \ifx #1\expandafter \@firstoftwo
 \else \expandafter \@secondoftwo
 \fi
}%
\providecommand \natexlab [1]{#1}%
\providecommand \enquote  [1]{``#1''}%
\providecommand \bibnamefont  [1]{#1}%
\providecommand \bibfnamefont [1]{#1}%
\providecommand \citenamefont [1]{#1}%
\providecommand \href@noop [0]{\@secondoftwo}%
\providecommand \href [0]{\begingroup \@sanitize@url \@href}%
\providecommand \@href[1]{\@@startlink{#1}\@@href}%
\providecommand \@@href[1]{\endgroup#1\@@endlink}%
\providecommand \@sanitize@url [0]{\catcode `\\12\catcode `\$12\catcode
  `\&12\catcode `\#12\catcode `\^12\catcode `\_12\catcode `\%12\relax}%
\providecommand \@@startlink[1]{}%
\providecommand \@@endlink[0]{}%
\providecommand \url  [0]{\begingroup\@sanitize@url \@url }%
\providecommand \@url [1]{\endgroup\@href {#1}{\urlprefix }}%
\providecommand \urlprefix  [0]{URL }%
\providecommand \Eprint [0]{\href }%
\providecommand \doibase [0]{http://dx.doi.org/}%
\providecommand \selectlanguage [0]{\@gobble}%
\providecommand \bibinfo  [0]{\@secondoftwo}%
\providecommand \bibfield  [0]{\@secondoftwo}%
\providecommand \translation [1]{[#1]}%
\providecommand \BibitemOpen [0]{}%
\providecommand \bibitemStop [0]{}%
\providecommand \bibitemNoStop [0]{.\EOS\space}%
\providecommand \EOS [0]{\spacefactor3000\relax}%
\providecommand \BibitemShut  [1]{\csname bibitem#1\endcsname}%
\let\auto@bib@innerbib\@empty
\bibitem [{\citenamefont {Faraday}(1831)}]{Faraday1831}%
  \BibitemOpen
  \bibfield  {author} {\bibinfo {author} {\bibfnamefont {M.}~\bibnamefont
  {Faraday}},\ }\href {\doibase 10.1098/rstl.1831.0018} {\bibfield  {journal}
  {\bibinfo  {journal} {Philosophical Transactions of the Royal Society of
  London}\ }\textbf {\bibinfo {volume} {121}},\ \bibinfo {pages} {299}
  (\bibinfo {year} {1831})}\BibitemShut {NoStop}%
\bibitem [{\citenamefont {Friend}\ and\ \citenamefont
  {Yeo}(2011)}]{Friend2011}%
  \BibitemOpen
  \bibfield  {author} {\bibinfo {author} {\bibfnamefont {J.}~\bibnamefont
  {Friend}}\ and\ \bibinfo {author} {\bibfnamefont {L.~Y.}\ \bibnamefont
  {Yeo}},\ }\href@noop {} {\bibfield  {journal} {\bibinfo  {journal} {Reviews
  of Modern Physics}\ }\textbf {\bibinfo {volume} {83}},\ \bibinfo {pages}
  {647} (\bibinfo {year} {2011})}\BibitemShut {NoStop}%
\bibitem [{\citenamefont {Sritharan}\ \emph {et~al.}(2006)\citenamefont
  {Sritharan}, \citenamefont {Strobl}, \citenamefont {Schneider},\ and\
  \citenamefont {Wixforth}}]{Sritharan2006}%
  \BibitemOpen
  \bibfield  {author} {\bibinfo {author} {\bibfnamefont {K.}~\bibnamefont
  {Sritharan}}, \bibinfo {author} {\bibfnamefont {C.~J.}\ \bibnamefont
  {Strobl}}, \bibinfo {author} {\bibfnamefont {M.~F.}\ \bibnamefont
  {Schneider}}, \ and\ \bibinfo {author} {\bibfnamefont {A.}~\bibnamefont
  {Wixforth}},\ }\href@noop {} {\bibfield  {journal} {\bibinfo  {journal}
  {Applied Physics Letters}\ }\textbf {\bibinfo {volume} {88}},\ \bibinfo
  {pages} {054102} (\bibinfo {year} {2006})}\BibitemShut {NoStop}%
\bibitem [{\citenamefont {Franke}\ \emph {et~al.}(2010)\citenamefont {Franke},
  \citenamefont {Braunmuller}, \citenamefont {Schmid}, \citenamefont
  {Wixforth},\ and\ \citenamefont {Weitz}}]{Franke2010}%
  \BibitemOpen
  \bibfield  {author} {\bibinfo {author} {\bibfnamefont {T.}~\bibnamefont
  {Franke}}, \bibinfo {author} {\bibfnamefont {S.}~\bibnamefont {Braunmuller}},
  \bibinfo {author} {\bibfnamefont {L.}~\bibnamefont {Schmid}}, \bibinfo
  {author} {\bibfnamefont {A.}~\bibnamefont {Wixforth}}, \ and\ \bibinfo
  {author} {\bibfnamefont {D.~A.}\ \bibnamefont {Weitz}},\ }\href@noop {}
  {\bibfield  {journal} {\bibinfo  {journal} {Lab on a Chip}\ }\textbf
  {\bibinfo {volume} {10}},\ \bibinfo {pages} {789} (\bibinfo {year}
  {2010})}\BibitemShut {NoStop}%
\bibitem [{\citenamefont {Lenshof}\ \emph {et~al.}(2012)\citenamefont
  {Lenshof}, \citenamefont {Magnusson},\ and\ \citenamefont
  {Laurell}}]{Lenshof2012}%
  \BibitemOpen
  \bibfield  {author} {\bibinfo {author} {\bibfnamefont {A.}~\bibnamefont
  {Lenshof}}, \bibinfo {author} {\bibfnamefont {C.}~\bibnamefont {Magnusson}},
  \ and\ \bibinfo {author} {\bibfnamefont {T.}~\bibnamefont {Laurell}},\
  }\href@noop {} {\bibfield  {journal} {\bibinfo  {journal} {Lab on a Chip}\
  }\textbf {\bibinfo {volume} {12}},\ \bibinfo {pages} {1210} (\bibinfo {year}
  {2012})}\BibitemShut {NoStop}%
\bibitem [{\citenamefont {Sadhal}(2012{\natexlab{a}})}]{Sadhal2012a}%
  \BibitemOpen
  \bibfield  {author} {\bibinfo {author} {\bibfnamefont {S.~S.}\ \bibnamefont
  {Sadhal}},\ }\href {\doibase 10.1039/c2lc40243b} {\bibfield  {journal}
  {\bibinfo  {journal} {Lab on a Chip}\ }\textbf {\bibinfo {volume} {12}},\
  \bibinfo {pages} {2600} (\bibinfo {year} {2012}{\natexlab{a}})}\BibitemShut
  {NoStop}%
\bibitem [{\citenamefont {Muller}\ \emph {et~al.}(2013)\citenamefont {Muller},
  \citenamefont {Rossi}, \citenamefont {Marin}, \citenamefont {Barnkop},
  \citenamefont {Augustsson}, \citenamefont {Laurell}, \citenamefont {Kahler},\
  and\ \citenamefont {Bruus}}]{Muller2013}%
  \BibitemOpen
  \bibfield  {author} {\bibinfo {author} {\bibfnamefont {P.~B.}\ \bibnamefont
  {Muller}}, \bibinfo {author} {\bibfnamefont {M.}~\bibnamefont {Rossi}},
  \bibinfo {author} {\bibfnamefont {A.~G.}\ \bibnamefont {Marin}}, \bibinfo
  {author} {\bibfnamefont {R.}~\bibnamefont {Barnkop}}, \bibinfo {author}
  {\bibfnamefont {P.}~\bibnamefont {Augustsson}}, \bibinfo {author}
  {\bibfnamefont {T.}~\bibnamefont {Laurell}}, \bibinfo {author} {\bibfnamefont
  {C.~J.}\ \bibnamefont {Kahler}}, \ and\ \bibinfo {author} {\bibfnamefont
  {H.}~\bibnamefont {Bruus}},\ }\href
  {https://journals.aps.org/pre/pdf/10.1103/PhysRevE.88.023006} {\bibfield
  {journal} {\bibinfo  {journal} {Physical Review E}\ }\textbf {\bibinfo
  {volume} {88}},\ \bibinfo {pages} {023006} (\bibinfo {year}
  {2013})}\BibitemShut {NoStop}%
\bibitem [{\citenamefont {Skov}\ \emph {et~al.}(2019)\citenamefont {Skov},
  \citenamefont {Sehgal}, \citenamefont {Kirby},\ and\ \citenamefont
  {Bruus}}]{Skov2019}%
  \BibitemOpen
  \bibfield  {author} {\bibinfo {author} {\bibfnamefont {N.~R.}\ \bibnamefont
  {Skov}}, \bibinfo {author} {\bibfnamefont {P.}~\bibnamefont {Sehgal}},
  \bibinfo {author} {\bibfnamefont {B.~J.}\ \bibnamefont {Kirby}}, \ and\
  \bibinfo {author} {\bibfnamefont {H.}~\bibnamefont {Bruus}},\ }\href
  {\doibase 10.1103/PhysRevApplied.12.044028} {\bibfield  {journal} {\bibinfo
  {journal} {Physical Review Applied}\ }\textbf {\bibinfo {volume} {12}},\
  \bibinfo {pages} {044028} (\bibinfo {year} {2019})}\BibitemShut {NoStop}%
\bibitem [{\citenamefont {Qiu}\ \emph {et~al.}(2019)\citenamefont {Qiu},
  \citenamefont {Karlsen}, \citenamefont {Bruus},\ and\ \citenamefont
  {Augustsson}}]{Qiu2019}%
  \BibitemOpen
  \bibfield  {author} {\bibinfo {author} {\bibfnamefont {W.}~\bibnamefont
  {Qiu}}, \bibinfo {author} {\bibfnamefont {J.~T.}\ \bibnamefont {Karlsen}},
  \bibinfo {author} {\bibfnamefont {H.}~\bibnamefont {Bruus}}, \ and\ \bibinfo
  {author} {\bibfnamefont {P.}~\bibnamefont {Augustsson}},\ }\href {\doibase
  10.1103/PhysRevApplied.11.024018} {\bibfield  {journal} {\bibinfo  {journal}
  {Physical Review Applied}\ }\textbf {\bibinfo {volume} {11}},\ \bibinfo
  {pages} {024018} (\bibinfo {year} {2019})}\BibitemShut {NoStop}%
\bibitem [{\citenamefont {Voth}\ \emph {et~al.}(2002)\citenamefont {Voth},
  \citenamefont {Bigger}, \citenamefont {Buckley}, \citenamefont {Losert},
  \citenamefont {Brenner}, \citenamefont {Stone},\ and\ \citenamefont
  {Gollub}}]{Voth2002}%
  \BibitemOpen
  \bibfield  {author} {\bibinfo {author} {\bibfnamefont {G.~A.}\ \bibnamefont
  {Voth}}, \bibinfo {author} {\bibfnamefont {B.}~\bibnamefont {Bigger}},
  \bibinfo {author} {\bibfnamefont {M.~R.}\ \bibnamefont {Buckley}}, \bibinfo
  {author} {\bibfnamefont {W.}~\bibnamefont {Losert}}, \bibinfo {author}
  {\bibfnamefont {M.~P.}\ \bibnamefont {Brenner}}, \bibinfo {author}
  {\bibfnamefont {H.~A.}\ \bibnamefont {Stone}}, \ and\ \bibinfo {author}
  {\bibfnamefont {J.~P.}\ \bibnamefont {Gollub}},\ }\href@noop {} {\bibfield
  {journal} {\bibinfo  {journal} {Physical Review Letters}\ }\textbf {\bibinfo
  {volume} {88}},\ \bibinfo {pages} {234301} (\bibinfo {year}
  {2002})}\BibitemShut {NoStop}%
\bibitem [{\citenamefont {Vuillermet}\ \emph {et~al.}(2016)\citenamefont
  {Vuillermet}, \citenamefont {Gires}, \citenamefont {Casset},\ and\
  \citenamefont {Poulain}}]{Vuillermet2016}%
  \BibitemOpen
  \bibfield  {author} {\bibinfo {author} {\bibfnamefont {G.}~\bibnamefont
  {Vuillermet}}, \bibinfo {author} {\bibfnamefont {P.-Y.}\ \bibnamefont
  {Gires}}, \bibinfo {author} {\bibfnamefont {F.}~\bibnamefont {Casset}}, \
  and\ \bibinfo {author} {\bibfnamefont {C.}~\bibnamefont {Poulain}},\
  }\href@noop {} {\bibfield  {journal} {\bibinfo  {journal} {Physical Review
  Letters}\ }\textbf {\bibinfo {volume} {116}},\ \bibinfo {pages} {184501}
  (\bibinfo {year} {2016})}\BibitemShut {NoStop}%
\bibitem [{\citenamefont {Legay}\ \emph {et~al.}(2012)\citenamefont {Legay},
  \citenamefont {Simony}, \citenamefont {Boldo}, \citenamefont {Gondrexon},
  \citenamefont {{Le Person}},\ and\ \citenamefont {Bontemps}}]{Legay2012}%
  \BibitemOpen
  \bibfield  {author} {\bibinfo {author} {\bibfnamefont {M.}~\bibnamefont
  {Legay}}, \bibinfo {author} {\bibfnamefont {B.}~\bibnamefont {Simony}},
  \bibinfo {author} {\bibfnamefont {P.}~\bibnamefont {Boldo}}, \bibinfo
  {author} {\bibfnamefont {N.}~\bibnamefont {Gondrexon}}, \bibinfo {author}
  {\bibfnamefont {S.}~\bibnamefont {{Le Person}}}, \ and\ \bibinfo {author}
  {\bibfnamefont {A.}~\bibnamefont {Bontemps}},\ }\href {\doibase
  10.1016/J.ULTSONCH.2012.04.001} {\bibfield  {journal} {\bibinfo  {journal}
  {Ultrasonics Sonochemistry}\ }\textbf {\bibinfo {volume} {19}},\ \bibinfo
  {pages} {1194} (\bibinfo {year} {2012})}\BibitemShut {NoStop}%
\bibitem [{\citenamefont {Loh}\ \emph {et~al.}(2002)\citenamefont {Loh},
  \citenamefont {Hyun}, \citenamefont {Ro},\ and\ \citenamefont
  {Kleinstreuer}}]{Loh2002}%
  \BibitemOpen
  \bibfield  {author} {\bibinfo {author} {\bibfnamefont {B.~G.}\ \bibnamefont
  {Loh}}, \bibinfo {author} {\bibfnamefont {S.}~\bibnamefont {Hyun}}, \bibinfo
  {author} {\bibfnamefont {P.~I.}\ \bibnamefont {Ro}}, \ and\ \bibinfo {author}
  {\bibfnamefont {C.}~\bibnamefont {Kleinstreuer}},\ }\href@noop {} {\bibfield
  {journal} {\bibinfo  {journal} {Journal of Acoustical Society of America}\
  }\textbf {\bibinfo {volume} {111}},\ \bibinfo {pages} {875} (\bibinfo {year}
  {2002})}\BibitemShut {NoStop}%
\bibitem [{\citenamefont {Westervelt}(1953)}]{Westervelt1953}%
  \BibitemOpen
  \bibfield  {author} {\bibinfo {author} {\bibfnamefont {P.~J.}\ \bibnamefont
  {Westervelt}},\ }\href {\doibase 10.1121/1.1907009} {\bibfield  {journal}
  {\bibinfo  {journal} {The Journal of the Acoustical Society of America}\
  }\textbf {\bibinfo {volume} {25}},\ \bibinfo {pages} {60} (\bibinfo {year}
  {1953})}\BibitemShut {NoStop}%
\bibitem [{\citenamefont {Nyborg}(1953)}]{Nyborg1953}%
  \BibitemOpen
  \bibfield  {author} {\bibinfo {author} {\bibfnamefont {W.~L.}\ \bibnamefont
  {Nyborg}},\ }\href {\doibase 10.1121/1.1907010} {\bibfield  {journal}
  {\bibinfo  {journal} {The Journal of the Acoustical Society of America}\
  }\textbf {\bibinfo {volume} {25}},\ \bibinfo {pages} {68} (\bibinfo {year}
  {1953})}\BibitemShut {NoStop}%
\bibitem [{\citenamefont {Lighthill}(1978)}]{Lighthill1978}%
  \BibitemOpen
  \bibfield  {author} {\bibinfo {author} {\bibfnamefont {S.~J.}\ \bibnamefont
  {Lighthill}},\ }\href@noop {} {\bibfield  {journal} {\bibinfo  {journal}
  {Journal of Sound And Vibration}\ }\textbf {\bibinfo {volume} {61}},\
  \bibinfo {pages} {391} (\bibinfo {year} {1978})}\BibitemShut {NoStop}%
\bibitem [{\citenamefont {Eckart}(1948)}]{Eckart1948}%
  \BibitemOpen
  \bibfield  {author} {\bibinfo {author} {\bibfnamefont {C.}~\bibnamefont
  {Eckart}},\ }\href {\doibase 10.1103/PhysRev.73.68} {\bibfield  {journal}
  {\bibinfo  {journal} {Physical Review}\ }\textbf {\bibinfo {volume} {73}},\
  \bibinfo {pages} {68} (\bibinfo {year} {1948})}\BibitemShut {NoStop}%
\bibitem [{\citenamefont {Rayleigh}(1884)}]{Rayleigh1884}%
  \BibitemOpen
  \bibfield  {author} {\bibinfo {author} {\bibfnamefont {L.}~\bibnamefont
  {Rayleigh}},\ }\href@noop {} {\bibfield  {journal} {\bibinfo  {journal}
  {Philosophical Transactions of the Royal Society of London}\ }\textbf
  {\bibinfo {volume} {175}},\ \bibinfo {pages} {1} (\bibinfo {year}
  {1884})}\BibitemShut {NoStop}%
\bibitem [{\citenamefont {Schlichting}\ and\ \citenamefont
  {Gersten}(2017)}]{Schlichting}%
  \BibitemOpen
  \bibfield  {author} {\bibinfo {author} {\bibfnamefont {H.}~\bibnamefont
  {Schlichting}}\ and\ \bibinfo {author} {\bibfnamefont {K.}~\bibnamefont
  {Gersten}},\ }\href@noop {} {\emph {\bibinfo {title} {Boundary-Layer
  Theory}}}\ (\bibinfo  {publisher} {Springer Nature},\ \bibinfo {year}
  {2017})\BibitemShut {NoStop}%
\bibitem [{\citenamefont {Nyborg}(1958)}]{Nyborg1958}%
  \BibitemOpen
  \bibfield  {author} {\bibinfo {author} {\bibfnamefont {W.~L.}\ \bibnamefont
  {Nyborg}},\ }\href {\doibase 10.1121/1.1909587} {\bibfield  {journal}
  {\bibinfo  {journal} {The Journal of the Acoustical Society of America}\
  }\textbf {\bibinfo {volume} {30}},\ \bibinfo {pages} {329} (\bibinfo {year}
  {1958})}\BibitemShut {NoStop}%
\bibitem [{\citenamefont {Riley}(1998)}]{Riley1998a}%
  \BibitemOpen
  \bibfield  {author} {\bibinfo {author} {\bibfnamefont {N.}~\bibnamefont
  {Riley}},\ }\href {\doibase 10.1007/s001620050068} {\emph {\bibinfo {title}
  {Theoretical and Computational Fluid Dynamics}}},\ Vol.~\bibinfo {volume}
  {10}\ (\bibinfo  {publisher} {Springer US},\ \bibinfo {address} {Boston,
  MA},\ \bibinfo {year} {1998})\ pp.\ \bibinfo {pages} {349--356}\BibitemShut
  {NoStop}%
\bibitem [{\citenamefont {Rayleigh}(2013)}]{Rayleigh2013}%
  \BibitemOpen
  \bibfield  {author} {\bibinfo {author} {\bibfnamefont {L.}~\bibnamefont
  {Rayleigh}},\ }\href {http://cds.cern.ch/record/105679} {\emph {\bibinfo
  {title} {{The Theory of Sound, Volume One.}}}}\ (\bibinfo  {publisher} {Dover
  Publications},\ \bibinfo {year} {2013})\ p.\ \bibinfo {pages}
  {985}\BibitemShut {NoStop}%
\bibitem [{\citenamefont {Da~Costa~Andrade}(1931)}]{Andrade1931}%
  \BibitemOpen
  \bibfield  {author} {\bibinfo {author} {\bibfnamefont {E.~N.}\ \bibnamefont
  {Da~Costa~Andrade}},\ }\href@noop {} {\bibfield  {journal} {\bibinfo
  {journal} {Proceedings of the Royal Society A}\ }\textbf {\bibinfo {volume}
  {134}},\ \bibinfo {pages} {445} (\bibinfo {year} {1931})}\BibitemShut
  {NoStop}%
\bibitem [{\citenamefont {Valverde}(2015)}]{Valverde2015}%
  \BibitemOpen
  \bibfield  {author} {\bibinfo {author} {\bibfnamefont {J.~M.}\ \bibnamefont
  {Valverde}},\ }\href@noop {} {\bibfield  {journal} {\bibinfo  {journal}
  {Contemporary Physics}\ }\textbf {\bibinfo {volume} {56}},\ \bibinfo {pages}
  {338} (\bibinfo {year} {2015})}\BibitemShut {NoStop}%
\bibitem [{\citenamefont {Hamilton}\ \emph {et~al.}(2002)\citenamefont
  {Hamilton}, \citenamefont {Ilinskii},\ and\ \citenamefont
  {Zabolotskaya}}]{Hamilton2002}%
  \BibitemOpen
  \bibfield  {author} {\bibinfo {author} {\bibfnamefont {M.~F.}\ \bibnamefont
  {Hamilton}}, \bibinfo {author} {\bibfnamefont {Y.~A.}\ \bibnamefont
  {Ilinskii}}, \ and\ \bibinfo {author} {\bibfnamefont {E.}~\bibnamefont
  {Zabolotskaya}},\ }\href@noop {} {\bibfield  {journal} {\bibinfo  {journal}
  {Journal of Acoustical Society of America}\ }\textbf {\bibinfo {volume}
  {113}},\ \bibinfo {pages} {153} (\bibinfo {year} {2002})}\BibitemShut
  {NoStop}%
\bibitem [{\citenamefont {Wiklund}\ \emph {et~al.}(2012)\citenamefont
  {Wiklund}, \citenamefont {Green},\ and\ \citenamefont
  {Ohlin}}]{Wiklund2012a}%
  \BibitemOpen
  \bibfield  {author} {\bibinfo {author} {\bibfnamefont {M.}~\bibnamefont
  {Wiklund}}, \bibinfo {author} {\bibfnamefont {R.}~\bibnamefont {Green}}, \
  and\ \bibinfo {author} {\bibfnamefont {M.}~\bibnamefont {Ohlin}},\ }\href
  {\doibase 10.1039/c2lc40203c} {\bibfield  {journal} {\bibinfo  {journal} {Lab
  on a Chip}\ }\textbf {\bibinfo {volume} {12}},\ \bibinfo {pages} {2438}
  (\bibinfo {year} {2012})}\BibitemShut {NoStop}%
\bibitem [{\citenamefont {Ahmed}\ \emph {et~al.}(2009)\citenamefont {Ahmed},
  \citenamefont {Mao}, \citenamefont {Juluri},\ and\ \citenamefont
  {Huang}}]{Ahmed2009}%
  \BibitemOpen
  \bibfield  {author} {\bibinfo {author} {\bibfnamefont {D.}~\bibnamefont
  {Ahmed}}, \bibinfo {author} {\bibfnamefont {X.}~\bibnamefont {Mao}}, \bibinfo
  {author} {\bibfnamefont {B.~K.}\ \bibnamefont {Juluri}}, \ and\ \bibinfo
  {author} {\bibfnamefont {T.~J.}\ \bibnamefont {Huang}},\ }\href@noop {}
  {\bibfield  {journal} {\bibinfo  {journal} {Microfluidics and Nanofluidics}\
  }\textbf {\bibinfo {volume} {7}},\ \bibinfo {pages} {727} (\bibinfo {year}
  {2009})}\BibitemShut {NoStop}%
\bibitem [{\citenamefont {Lu}\ \emph {et~al.}(2019)\citenamefont {Lu},
  \citenamefont {Zhao}, \citenamefont {Peng}, \citenamefont {Li},\ and\
  \citenamefont {Liu}}]{Lux2019}%
  \BibitemOpen
  \bibfield  {author} {\bibinfo {author} {\bibfnamefont {X.}~\bibnamefont
  {Lu}}, \bibinfo {author} {\bibfnamefont {K.}~\bibnamefont {Zhao}}, \bibinfo
  {author} {\bibfnamefont {H.}~\bibnamefont {Peng}}, \bibinfo {author}
  {\bibfnamefont {H.}~\bibnamefont {Li}}, \ and\ \bibinfo {author}
  {\bibfnamefont {W.}~\bibnamefont {Liu}},\ }\href {\doibase
  10.1103/PhysRevApplied.11.044064} {\bibfield  {journal} {\bibinfo  {journal}
  {Physical Review Applied}\ }\textbf {\bibinfo {volume} {11}},\ \bibinfo
  {pages} {1} (\bibinfo {year} {2019})}\BibitemShut {NoStop}%
\bibitem [{\citenamefont {Lei}\ \emph {et~al.}(2018)\citenamefont {Lei},
  \citenamefont {Hill}, \citenamefont {de~Le{\'o}n~Albarr{\'a}n},\ and\
  \citenamefont {Glynne‑Jones}}]{Lei2018}%
  \BibitemOpen
  \bibfield  {author} {\bibinfo {author} {\bibfnamefont {J.}~\bibnamefont
  {Lei}}, \bibinfo {author} {\bibfnamefont {M.}~\bibnamefont {Hill}}, \bibinfo
  {author} {\bibfnamefont {C.~P.}\ \bibnamefont {de~Le{\'o}n~Albarr{\'a}n}}, \
  and\ \bibinfo {author} {\bibfnamefont {P.}~\bibnamefont {Glynne‑Jones}},\
  }\href@noop {} {\bibfield  {journal} {\bibinfo  {journal} {Microfluidics and
  Nanofluidics}\ }\textbf {\bibinfo {volume} {22}},\ \bibinfo {pages} {140}
  (\bibinfo {year} {2018})}\BibitemShut {NoStop}%
\bibitem [{\citenamefont {Subbotin}\ \emph {et~al.}(2019)\citenamefont
  {Subbotin}, \citenamefont {Kozlov},\ and\ \citenamefont
  {Shiryaeva}}]{Subbotin2019}%
  \BibitemOpen
  \bibfield  {author} {\bibinfo {author} {\bibfnamefont {S.}~\bibnamefont
  {Subbotin}}, \bibinfo {author} {\bibfnamefont {V.}~\bibnamefont {Kozlov}}, \
  and\ \bibinfo {author} {\bibfnamefont {M.}~\bibnamefont {Shiryaeva}},\
  }\href@noop {} {\bibfield  {journal} {\bibinfo  {journal} {Phys. Fluids}\
  }\textbf {\bibinfo {volume} {31}},\ \bibinfo {pages} {103604} (\bibinfo
  {year} {2019})}\BibitemShut {NoStop}%
\bibitem [{\citenamefont {Jannesar}\ and\ \citenamefont
  {Hamzehpour}(2019)}]{Jannesar2019}%
  \BibitemOpen
  \bibfield  {author} {\bibinfo {author} {\bibfnamefont {E.~A.}\ \bibnamefont
  {Jannesar}}\ and\ \bibinfo {author} {\bibfnamefont {H.}~\bibnamefont
  {Hamzehpour}},\ }\href@noop {} {\bibfield  {journal} {\bibinfo  {journal}
  {arXiv}\ } (\bibinfo {year} {2019})},\ \Eprint
  {http://arxiv.org/abs/1909.03251v1 [physics.flu-dyn]} {1909.03251v1
  [physics.flu-dyn]} \BibitemShut {NoStop}%
\bibitem [{\citenamefont {Lei}\ \emph {et~al.}(2017{\natexlab{a}})\citenamefont
  {Lei}, \citenamefont {Hill},\ and\ \citenamefont {Glynne-Jones}}]{Lei2017b}%
  \BibitemOpen
  \bibfield  {author} {\bibinfo {author} {\bibfnamefont {J.}~\bibnamefont
  {Lei}}, \bibinfo {author} {\bibfnamefont {M.}~\bibnamefont {Hill}}, \ and\
  \bibinfo {author} {\bibfnamefont {P.}~\bibnamefont {Glynne-Jones}},\
  }\href@noop {} {\bibfield  {journal} {\bibinfo  {journal} {Physical Review
  Applied}\ }\textbf {\bibinfo {volume} {8}},\ \bibinfo {pages} {014018}
  (\bibinfo {year} {2017}{\natexlab{a}})}\BibitemShut {NoStop}%
\bibitem [{\citenamefont {Huang}\ \emph {et~al.}(2013)\citenamefont {Huang},
  \citenamefont {Xie}, \citenamefont {Ahmed}, \citenamefont {Rufo},
  \citenamefont {Nama}, \citenamefont {Chen}, \citenamefont {Chan},\ and\
  \citenamefont {Huang}}]{Huang2013a}%
  \BibitemOpen
  \bibfield  {author} {\bibinfo {author} {\bibfnamefont {P.~H.}\ \bibnamefont
  {Huang}}, \bibinfo {author} {\bibfnamefont {Y.}~\bibnamefont {Xie}}, \bibinfo
  {author} {\bibfnamefont {D.}~\bibnamefont {Ahmed}}, \bibinfo {author}
  {\bibfnamefont {J.}~\bibnamefont {Rufo}}, \bibinfo {author} {\bibfnamefont
  {N.}~\bibnamefont {Nama}}, \bibinfo {author} {\bibfnamefont {Y.}~\bibnamefont
  {Chen}}, \bibinfo {author} {\bibfnamefont {C.~Y.}\ \bibnamefont {Chan}}, \
  and\ \bibinfo {author} {\bibfnamefont {T.~J.}\ \bibnamefont {Huang}},\ }\href
  {\doibase 10.1039/c3lc50568e} {\bibfield  {journal} {\bibinfo  {journal} {Lab
  on a Chip}\ }\textbf {\bibinfo {volume} {13}},\ \bibinfo {pages} {3847}
  (\bibinfo {year} {2013})}\BibitemShut {NoStop}%
\bibitem [{\citenamefont {Huang}\ \emph {et~al.}(2014)\citenamefont {Huang},
  \citenamefont {Nama}, \citenamefont {Mao}, \citenamefont {Li}, \citenamefont
  {Rufo}, \citenamefont {Chen}, \citenamefont {Xie}, \citenamefont {Wei},
  \citenamefont {Wang},\ and\ \citenamefont {Huang}}]{Huang2014}%
  \BibitemOpen
  \bibfield  {author} {\bibinfo {author} {\bibfnamefont {P.~H.}\ \bibnamefont
  {Huang}}, \bibinfo {author} {\bibfnamefont {N.}~\bibnamefont {Nama}},
  \bibinfo {author} {\bibfnamefont {Z.}~\bibnamefont {Mao}}, \bibinfo {author}
  {\bibfnamefont {P.}~\bibnamefont {Li}}, \bibinfo {author} {\bibfnamefont
  {J.}~\bibnamefont {Rufo}}, \bibinfo {author} {\bibfnamefont {Y.}~\bibnamefont
  {Chen}}, \bibinfo {author} {\bibfnamefont {Y.}~\bibnamefont {Xie}}, \bibinfo
  {author} {\bibfnamefont {C.~H.}\ \bibnamefont {Wei}}, \bibinfo {author}
  {\bibfnamefont {L.}~\bibnamefont {Wang}}, \ and\ \bibinfo {author}
  {\bibfnamefont {T.~J.}\ \bibnamefont {Huang}},\ }\href
  {http://xlink.rsc.org/?DOI=C4LC00806E} {\bibfield  {journal} {\bibinfo
  {journal} {Lab on a Chip}\ }\textbf {\bibinfo {volume} {14}},\ \bibinfo
  {pages} {4319} (\bibinfo {year} {2014})}\BibitemShut {NoStop}%
\bibitem [{\citenamefont {Nama}\ \emph {et~al.}(2014)\citenamefont {Nama},
  \citenamefont {Huang}, \citenamefont {Huang},\ and\ \citenamefont
  {Costanzo}}]{Nama2014}%
  \BibitemOpen
  \bibfield  {author} {\bibinfo {author} {\bibfnamefont {N.}~\bibnamefont
  {Nama}}, \bibinfo {author} {\bibfnamefont {P.~H.}\ \bibnamefont {Huang}},
  \bibinfo {author} {\bibfnamefont {T.~J.}\ \bibnamefont {Huang}}, \ and\
  \bibinfo {author} {\bibfnamefont {F.}~\bibnamefont {Costanzo}},\ }\href
  {http://xlink.rsc.org/?DOI=C4LC00191E} {\bibfield  {journal} {\bibinfo
  {journal} {Lab on a Chip}\ }\textbf {\bibinfo {volume} {14}},\ \bibinfo
  {pages} {2824} (\bibinfo {year} {2014})}\BibitemShut {NoStop}%
\bibitem [{\citenamefont {Nama}\ \emph {et~al.}(2016)\citenamefont {Nama},
  \citenamefont {Huang}, \citenamefont {Huang},\ and\ \citenamefont
  {Costanzo}}]{Nama2016a}%
  \BibitemOpen
  \bibfield  {author} {\bibinfo {author} {\bibfnamefont {N.}~\bibnamefont
  {Nama}}, \bibinfo {author} {\bibfnamefont {P.~H.}\ \bibnamefont {Huang}},
  \bibinfo {author} {\bibfnamefont {T.~J.}\ \bibnamefont {Huang}}, \ and\
  \bibinfo {author} {\bibfnamefont {F.}~\bibnamefont {Costanzo}},\ }\href
  {http://dx.doi.org/10.1063/1.4946875} {\bibfield  {journal} {\bibinfo
  {journal} {Biomicrofluidics}\ }\textbf {\bibinfo {volume} {10}},\ \bibinfo
  {pages} {024124} (\bibinfo {year} {2016})}\BibitemShut {NoStop}%
\bibitem [{\citenamefont {Zhang}\ \emph {et~al.}(2019)\citenamefont {Zhang},
  \citenamefont {Guo}, \citenamefont {Brunet}, \citenamefont {Costalonga},\
  and\ \citenamefont {Royon}}]{Zhang2019}%
  \BibitemOpen
  \bibfield  {author} {\bibinfo {author} {\bibfnamefont {C.}~\bibnamefont
  {Zhang}}, \bibinfo {author} {\bibfnamefont {X.}~\bibnamefont {Guo}}, \bibinfo
  {author} {\bibfnamefont {P.}~\bibnamefont {Brunet}}, \bibinfo {author}
  {\bibfnamefont {M.}~\bibnamefont {Costalonga}}, \ and\ \bibinfo {author}
  {\bibfnamefont {L.}~\bibnamefont {Royon}},\ }\href {\doibase
  10.1007/s10404-019-2271-5} {\bibfield  {journal} {\bibinfo  {journal}
  {Microfluidics and Nanofluidics}\ }\textbf {\bibinfo {volume} {23}},\
  \bibinfo {pages} {104} (\bibinfo {year} {2019})}\BibitemShut {NoStop}%
\bibitem [{\citenamefont {Huang}\ \emph {et~al.}(2018)\citenamefont {Huang},
  \citenamefont {Chan}, \citenamefont {Li}, \citenamefont {Wang}, \citenamefont
  {Nama}, \citenamefont {Bachman},\ and\ \citenamefont {Huang}}]{Huang2018a}%
  \BibitemOpen
  \bibfield  {author} {\bibinfo {author} {\bibfnamefont {P.~H.}\ \bibnamefont
  {Huang}}, \bibinfo {author} {\bibfnamefont {C.~Y.}\ \bibnamefont {Chan}},
  \bibinfo {author} {\bibfnamefont {P.}~\bibnamefont {Li}}, \bibinfo {author}
  {\bibfnamefont {Y.}~\bibnamefont {Wang}}, \bibinfo {author} {\bibfnamefont
  {N.}~\bibnamefont {Nama}}, \bibinfo {author} {\bibfnamefont {H.}~\bibnamefont
  {Bachman}}, \ and\ \bibinfo {author} {\bibfnamefont {T.~J.}\ \bibnamefont
  {Huang}},\ }\href {\doibase 10.1039/C8LC00193F} {\bibfield  {journal}
  {\bibinfo  {journal} {Lab on a Chip}\ }\textbf {\bibinfo {volume} {18}},\
  \bibinfo {pages} {1411} (\bibinfo {year} {2018})}\BibitemShut {NoStop}%
\bibitem [{\citenamefont {Leibacher}\ \emph {et~al.}(2015)\citenamefont
  {Leibacher}, \citenamefont {Hahn},\ and\ \citenamefont
  {Dual}}]{Leibacher2015}%
  \BibitemOpen
  \bibfield  {author} {\bibinfo {author} {\bibfnamefont {I.}~\bibnamefont
  {Leibacher}}, \bibinfo {author} {\bibfnamefont {P.}~\bibnamefont {Hahn}}, \
  and\ \bibinfo {author} {\bibfnamefont {J.}~\bibnamefont {Dual}},\ }\href
  {\doibase 10.1007/s10404-015-1621-1} {\bibfield  {journal} {\bibinfo
  {journal} {Microfluidics and Nanofluidics}\ }\textbf {\bibinfo {volume}
  {19}},\ \bibinfo {pages} {923} (\bibinfo {year} {2015})}\BibitemShut
  {NoStop}%
\bibitem [{\citenamefont {Cao}\ and\ \citenamefont {Lu}(2016)}]{Cao2016}%
  \BibitemOpen
  \bibfield  {author} {\bibinfo {author} {\bibfnamefont {Z.}~\bibnamefont
  {Cao}}\ and\ \bibinfo {author} {\bibfnamefont {C.}~\bibnamefont {Lu}},\
  }\href {\doibase 10.1021/acs.analchem.5b04707} {\bibfield  {journal}
  {\bibinfo  {journal} {Analytical Chemistry}\ }\textbf {\bibinfo {volume}
  {88}},\ \bibinfo {pages} {1965} (\bibinfo {year} {2016})}\BibitemShut
  {NoStop}%
\bibitem [{\citenamefont {Bachman}\ \emph {et~al.}(2018)\citenamefont
  {Bachman}, \citenamefont {Huang}, \citenamefont {Zhao}, \citenamefont {Yang},
  \citenamefont {Zhang}, \citenamefont {Fu},\ and\ \citenamefont
  {Huang}}]{Bachman2018}%
  \BibitemOpen
  \bibfield  {author} {\bibinfo {author} {\bibfnamefont {H.}~\bibnamefont
  {Bachman}}, \bibinfo {author} {\bibfnamefont {P.~H.}\ \bibnamefont {Huang}},
  \bibinfo {author} {\bibfnamefont {S.}~\bibnamefont {Zhao}}, \bibinfo {author}
  {\bibfnamefont {S.}~\bibnamefont {Yang}}, \bibinfo {author} {\bibfnamefont
  {P.}~\bibnamefont {Zhang}}, \bibinfo {author} {\bibfnamefont
  {H.}~\bibnamefont {Fu}}, \ and\ \bibinfo {author} {\bibfnamefont {T.~J.}\
  \bibnamefont {Huang}},\ }\href {\doibase 10.1039/C7LC01222E} {\bibfield
  {journal} {\bibinfo  {journal} {Lab on a Chip}\ }\textbf {\bibinfo {volume}
  {18}},\ \bibinfo {pages} {433} (\bibinfo {year} {2018})}\BibitemShut
  {NoStop}%
\bibitem [{\citenamefont {Ovchinnikov}\ \emph {et~al.}(2014)\citenamefont
  {Ovchinnikov}, \citenamefont {Zhou},\ and\ \citenamefont
  {Yalamanchili}}]{Ovchinnikov2014}%
  \BibitemOpen
  \bibfield  {author} {\bibinfo {author} {\bibfnamefont {M.}~\bibnamefont
  {Ovchinnikov}}, \bibinfo {author} {\bibfnamefont {J.}~\bibnamefont {Zhou}}, \
  and\ \bibinfo {author} {\bibfnamefont {S.}~\bibnamefont {Yalamanchili}},\
  }\href {\doibase 10.1121/1.4881919} {\bibfield  {journal} {\bibinfo
  {journal} {The Journal of the Acoustical Society of America}\ }\textbf
  {\bibinfo {volume} {136}},\ \bibinfo {pages} {22} (\bibinfo {year}
  {2014})}\BibitemShut {NoStop}%
\bibitem [{\citenamefont {Bruus}(2012)}]{Bruus2012d}%
  \BibitemOpen
  \bibfield  {author} {\bibinfo {author} {\bibfnamefont {H.}~\bibnamefont
  {Bruus}},\ }\href {\doibase 10.1039/C1LC20770A} {\bibfield  {journal}
  {\bibinfo  {journal} {Lab on a Chip}\ }\textbf {\bibinfo {volume} {12}},\
  \bibinfo {pages} {20} (\bibinfo {year} {2012})}\BibitemShut {NoStop}%
\bibitem [{\citenamefont {Sadhal}(2012{\natexlab{b}})}]{Sadhal2012}%
  \BibitemOpen
  \bibfield  {author} {\bibinfo {author} {\bibfnamefont {S.}~\bibnamefont
  {Sadhal}},\ }\href@noop {} {\bibfield  {journal} {\bibinfo  {journal} {Lab on
  a Chip}\ }\textbf {\bibinfo {volume} {12}},\ \bibinfo {pages} {2292}
  (\bibinfo {year} {2012}{\natexlab{b}})}\BibitemShut {NoStop}%
\bibitem [{\citenamefont {Boluriaan}\ and\ \citenamefont
  {Morris}(2003)}]{Boluriaan2003a}%
  \BibitemOpen
  \bibfield  {author} {\bibinfo {author} {\bibfnamefont {S.}~\bibnamefont
  {Boluriaan}}\ and\ \bibinfo {author} {\bibfnamefont {P.}~\bibnamefont
  {Morris}},\ }\href {\doibase 10.1260/147547203322986142} {\bibfield
  {journal} {\bibinfo  {journal} {International Journal of Aeroacoustics}\
  }\textbf {\bibinfo {volume} {2}},\ \bibinfo {pages} {255} (\bibinfo {year}
  {2003})}\BibitemShut {NoStop}%
\bibitem [{\citenamefont {Lei}\ \emph {et~al.}(2017{\natexlab{b}})\citenamefont
  {Lei}, \citenamefont {Glynne-Jones},\ and\ \citenamefont {Hill}}]{Lei2017}%
  \BibitemOpen
  \bibfield  {author} {\bibinfo {author} {\bibfnamefont {J.}~\bibnamefont
  {Lei}}, \bibinfo {author} {\bibfnamefont {P.}~\bibnamefont {Glynne-Jones}}, \
  and\ \bibinfo {author} {\bibfnamefont {M.}~\bibnamefont {Hill}},\ }\href
  {\doibase 10.1007/s10404-017-1865-z} {\bibfield  {journal} {\bibinfo
  {journal} {Microfluidics and Nanofluidics}\ }\textbf {\bibinfo {volume}
  {21}},\ \bibinfo {pages} {23} (\bibinfo {year}
  {2017}{\natexlab{b}})}\BibitemShut {NoStop}%
\bibitem [{Com()}]{Comsol}%
  \BibitemOpen
  \href@noop {} {\emph {\bibinfo {title} {COMSOL Multiphysics{\textregistered}
  v. 5.3}}},\ \bibinfo {organization} {www.comsol.com},\ \bibinfo {address}
  {COMSOL AB, Stockholm, Sweden}\BibitemShut {NoStop}%
\bibitem [{\citenamefont {Costalonga}\ \emph {et~al.}(2015)\citenamefont
  {Costalonga}, \citenamefont {Brunet},\ and\ \citenamefont
  {Peerhossaini}}]{Costalonga2015}%
  \BibitemOpen
  \bibfield  {author} {\bibinfo {author} {\bibfnamefont {M.}~\bibnamefont
  {Costalonga}}, \bibinfo {author} {\bibfnamefont {P.}~\bibnamefont {Brunet}},
  \ and\ \bibinfo {author} {\bibfnamefont {H.}~\bibnamefont {Peerhossaini}},\
  }\href@noop {} {\bibfield  {journal} {\bibinfo  {journal} {Physics of
  Fluids}\ }\textbf {\bibinfo {volume} {27}},\ \bibinfo {pages} {013101}
  (\bibinfo {year} {2015})}\BibitemShut {NoStop}%
\bibitem [{\citenamefont {Bahrani}\ \emph {et~al.}(2020)\citenamefont
  {Bahrani}, \citenamefont {P{\'e}rinet}, \citenamefont {Costalonga},
  \citenamefont {Royon},\ and\ \citenamefont {Brunet}}]{bahrani2020vortex}%
  \BibitemOpen
  \bibfield  {author} {\bibinfo {author} {\bibfnamefont {S.~A.}\ \bibnamefont
  {Bahrani}}, \bibinfo {author} {\bibfnamefont {N.}~\bibnamefont
  {P{\'e}rinet}}, \bibinfo {author} {\bibfnamefont {M.}~\bibnamefont
  {Costalonga}}, \bibinfo {author} {\bibfnamefont {L.}~\bibnamefont {Royon}}, \
  and\ \bibinfo {author} {\bibfnamefont {P.}~\bibnamefont {Brunet}},\
  }\href@noop {} {\bibfield  {journal} {\bibinfo  {journal} {arXiv}\ }\textbf
  {\bibinfo {volume} {2001}},\ \bibinfo {pages} {01131} (\bibinfo {year}
  {2020})}\BibitemShut {NoStop}%
\bibitem [{\citenamefont {Guo}\ \emph {et~al.}(2013)\citenamefont {Guo},
  \citenamefont {Fan},\ and\ \citenamefont {Luo}}]{Guo2013}%
  \BibitemOpen
  \bibfield  {author} {\bibinfo {author} {\bibfnamefont {X.}~\bibnamefont
  {Guo}}, \bibinfo {author} {\bibfnamefont {Y.}~\bibnamefont {Fan}}, \ and\
  \bibinfo {author} {\bibfnamefont {L.}~\bibnamefont {Luo}},\ }\href {\doibase
  10.1016/j.cej.2012.08.068} {\bibfield  {journal} {\bibinfo  {journal}
  {Chemical Engineering Journal}\ }\textbf {\bibinfo {volume} {227}},\ \bibinfo
  {pages} {116} (\bibinfo {year} {2013})}\BibitemShut {NoStop}%
\bibitem [{\citenamefont {Muller}\ and\ \citenamefont
  {Bruus}(2015)}]{Muller2015}%
  \BibitemOpen
  \bibfield  {author} {\bibinfo {author} {\bibfnamefont {P.~B.}\ \bibnamefont
  {Muller}}\ and\ \bibinfo {author} {\bibfnamefont {H.}~\bibnamefont {Bruus}},\
  }\href {https://doi.org/10.1103/physreve.92.063018} {\bibfield  {journal}
  {\bibinfo  {journal} {Physical Review E - Statistical, Nonlinear, and Soft
  Matter Physics}\ }\textbf {\bibinfo {volume} {92}} (\bibinfo {year}
  {2015})}\BibitemShut {NoStop}%
\end{thebibliography}%

\end{document}